# Getting more from your regression model: A free lunch?

**David P. Hofmeyr**
**Department of Statistics and Actuarial Science, Stellenbosch University**

## Abstract

We consider a simple approach for approximating detailed information about the conditional distribution of a real-valued response variable, given values for its covariates, using only the outputs from a standard regression model. We validate this approach by assessing its performance in the context of quantile regression; when applied to the outputs of linear, gradient boosted tree ensemble and random forest models. We find that it compares favourably to the standard approach for estimating quantile regression functions, especially for commonly selected tail probabilities, and is highly competitive with the quantile regression forest model, across a large collection of benchmark data sets.

# 1 Introduction

The problem of regression is associated with estimating properties of the conditional distribution of the *response variable*, $Y$, given values for the *covariates*, represented as the random vector $X = (X_1, ..., X_p)$. By far the most common regression problem focuses on estimating the conditional expectation, which we will denote $\mu(\mathbf{x}) := E[Y|X = \mathbf{x}]$, where the modelling assumption on the joint distribution of $Y$ and $X$ is expressed through the *regression equation*,

$$Y = \mu(X) + \epsilon, \tag{1}$$

where $\epsilon$ is a *residual* term satisfying $E[\epsilon|X = \mathbf{x}] = 0\ \forall\mathbf{x}$, and which, in the simplest context, is assumed to be independent of $X$ and whose distribution, or distribution family, may be assumed known. In practice we make observations of pairs of $Y$ and $X$, say $(y_1, \mathbf{x}_1), ..., (y_n, \mathbf{x}_n)$, assumed to have been drawn independently from their joint distribution. It is well known that the function which minimises the *expected squared error*, $\text{argmin}_g E[(Y - g(X))^2]$, is the function $\mu$, and this motivates estimating $\mu$ using the mean-squared-error (MSE) criterion, i.e., by setting

$$\hat{\mu} = \text{argmin}_{g \in \mathcal{F}} \frac{1}{n} \sum_{i=1}^{n} (y_i - g(\mathbf{x}_i))^2, \tag{2}$$

where $\mathcal{F}$ is a chosen class of functions from which to select ones model. With such an estimate for $\mu$, one is able to obtain a point estimate for $E[Y|X = \mathbf{x}]$ for any *query point*, $\mathbf{x}$, by simply taking the value $\hat{\mu}(\mathbf{x})$. Such point estimates are frequently used as *predictions* for what an unobserved response value associated with this query point might be.

When more detail about the conditional distribution of $Y|X = \mathbf{x}$, than just the mean, is desired, quantile regression may be used to estimate the function

$$q_\alpha(\mathbf{x}) := \inf\{y \in \mathbb{R} | F_{Y|X=\mathbf{x}}(y) \geq \alpha\},$$

for any $\alpha \in (0, 1)$, where $F_{Y|X=\mathbf{x}}$ is the conditional distribution function of $Y|X = \mathbf{x}$. In this context it is known that

$$q_\alpha = \text{argmin}_g E\left[(1 - \alpha)(g(X) - Y)_+ + \alpha(Y - g(X))_+\right], \tag{3}$$

where $(z)_+ = \max\{z, 0\}$ for $z \in \mathbb{R}$. The standard approach to conditional quantile estimation, or *quantile regression*, therefore focuses on minimising the *sample quantile error*, i.e.,

$$\hat{q}_\alpha = \text{argmin}_{g \in \mathcal{F}} \frac{1}{n} \sum_{i=1}^{n} \left( (1 - \alpha) \sum_{i: y_i < g(\mathbf{x}_i)} (g(\mathbf{x}_i) - y_i) + \alpha \sum_{i: y_i \geq g(\mathbf{x}_i)} (y_i - g(\mathbf{x}_i)) \right). \tag{4}$$

One of the useful features of these quantile regression functions is that they satisfy

$$P(q_\alpha(\mathbf{x}) < Y \leq q_{1-\alpha}(\mathbf{x}) | X = \mathbf{x}) = 1 - 2\alpha,$$

and so estimates for these functions allow one to obtain *prediction intervals* around ones point estimates which, provided they are accurately estimated, indicate an interval which contains an unobserved response value associated with a given query point with a desired probability.

Note also that, in principle, the full conditional distribution of the response can be estimated to chosen precision with enough of these quantile functions. However, the computational burden which such a task may represent when using the standard approach of minimising (4) notwithstanding, another possible issue with this is that when estimates for $\{q_\alpha\}_{\alpha \in (0,1)}$ are selected from very rich classes of functions, the associated optimisation problems tend to be non-convex. As a result the implicit ordering $\alpha < \alpha' \Rightarrow q_\alpha(\mathbf{x}) \leq q_{\alpha'}(\mathbf{x}) \ \forall \mathbf{x}$ may, in theory, not hold for the estimates.

In this paper we discuss a simple and computationally efficient approach for approximating (potentially very numerous) quantile functions which is guaranteed to preserve ordering, and which requires only a suitably chosen estimate for the mean function, $\mu$. The approach is based on the recognition that in many application areas the distribution of $Y|X$ is largely determined by its mean, i.e., that $Y|X \overset{D}{\approx} Y|\mu(X)$. When the function $\mu$ is one-to-one, this equivalence holds exactly, and in the presence of boundary and threshold effects on the response, and, for example, when the coefficient of variation is constant in $Y$ w.r.t. $X$, it will often hold approximately. In fact, as we find, conditional quantile estimation arising from simple non-parametric smoothing and relying on the assumption that this approximation holds is reliable across a wide range of applications. In particular, we test its effectiveness on 50 benchmark data sets taken from the `R` package `pmlbr` (Le et al., 2020) and find that it outperforms quantile regression forests (Meinshausen and Ridgeway, 2006) and quantile estimation arising by minimising (4) in the majority of the cases.

In the following section we give an overview of our proposed method. We also provide high level discussion on some of its strengths and limitations and illustrate some of these by way of a simulated example. In Section 3 we provide results from an extensive set of experiments motivated by investigating the performance of the proposed method in comparison with existing alternatives. We give a concluding discussion and some recommendations in Section 4.

## 2 Conditional estimation from weighted averages, determined from the outputs of a standard regression model

Suppose we are interested in estimating some property of the conditional distribution of $Y|X = \mathbf{x}$ which is expressible as $\kappa(\mathbf{x}) := E[k(Y)|X = \mathbf{x}]$, for some function $k$. At the essence of the standard non-parametric smoothing approach is, somewhat loosely, the process of sorting the observations of $Y$ into those which are more and less *relevant* for the task, and then defining $\hat{\kappa}(\mathbf{x})$ as a weighted average of the quantities $k(y_i); i = 1, ..., n$, in which more relevant points are given greater weights than less relevant ones. The standard assumption is that $\kappa(\mathbf{x})$ varies smoothly in $\mathbf{x}$, and so $\mathbf{x} \approx \mathbf{x}' \Rightarrow \kappa(\mathbf{x}) \approx \kappa(\mathbf{x}')$. Any observation, $y_i$, for which the corresponding $\mathbf{x}_i$ is close to a point of interest/query point, $\mathbf{x}$, can then be seen as relevant for estimating $\kappa(\mathbf{x})$, since $k(y_i)$ is a realisation from a distribution with mean close to $\kappa(\mathbf{x})$. Those $y_i$'s whose corresponding $\mathbf{x}_i$'s are nearer $\mathbf{x}$ are therefore given a higher weight than those whose are more distant. Provided the weights assigned to points outside a neighbourhood of $\mathbf{x}$ which shrinks appropriately fast as the sample size increases, this approach achieves consistent estimation.

Very simply then, whenever the distribution of $Y|X = \mathbf{x}$ can be approximated by that of $Y|\mu(X) = \mu(\mathbf{x})$, at least over a large part of the support of $X$, the conditions under which a point is relevant for estimation at the query point, $\mathbf{x}$, may be replaced by $\mu(\mathbf{x}_i) \approx \mu(\mathbf{x})$, or, in a practical context, by $\hat{\mu}(\mathbf{x}_i) \approx \hat{\mu}(\mathbf{x})$, where $\hat{\mu}$ is an appropriately selected estimate for $\mu$. We may therefore consider estimates of the form

$$\hat{\kappa}(\mathbf{x}) := \sum_{i=1}^{n} w_i(\mathbf{x})k(y_i),$$

where the *weight functions*, $w_i; i = 1, ..., n$, are non-negative and satisfy $\sum_{i=1}^{n} w_i(\mathbf{x}) = 1 \ \forall \mathbf{x}$, and are monotonically decreasing in the quantities $|\hat{\mu}(\mathbf{x}_i) - \hat{\mu}(\mathbf{x})|$. In particular, we use a kernel to determine these weights, and so set

$$w_i(\mathbf{x}) = \frac{K\left(\frac{\hat{\mu}(\mathbf{x}_i)-\hat{\mu}(\mathbf{x})}{h}\right)}{\sum_{j=1}^n K\left(\frac{\hat{\mu}(\mathbf{x}_j)-\hat{\mu}(\mathbf{x})}{h}\right)}, \tag{5}$$

where $K : \mathbb{R} \to \mathbb{R}^+$ is the kernel and $h > 0$ is an appropriate smoothing parameter, referred to as the *bandwidth*. The result is that $\hat{\kappa}(\mathbf{x})$ is equivalent to a Nadaraya-Watson estimate (Nadaraya, 1964; Watson, 1964) obtained from regressing the $k(y_i)$'s on the outputs from $\hat{\mu}$.

**Remark 1** *A useful way of thinking about how the assumption that* $Y|X = \mathbf{x} \overset{D}{\approx} Y|\mu(X) = \mu(\mathbf{x})$ *affects estimation at* $\mathbf{x}$ *is to recognise that, at least at a high level, the estimation of* $\kappa(\mathbf{x})$ *is based on a sample from a mixture distribution in which the component distributions are given by the set* $\{F_{Y|X=\mathbf{x}'}|\hat{\mu}(\mathbf{x}') \approx \hat{\mu}(\mathbf{x})\}$ *and the mixing proportions are approximately proportional to the elements in the set* $\{f_X(\mathbf{x}')|\hat{\mu}(\mathbf{x}') \approx \hat{\mu}(\mathbf{x})\}$*, where* $f_X$ *is the density of* $X$*.*

## 2.1 Quantile estimates from weighted averages

Before continuing we will briefly explain how conditional quantiles can be estimated within the proposed framework. Although conditional quantiles cannot be directly expressed in the form $E[k(Y)|X = \mathbf{x}]$, it is possible to use this formulation to estimate the conditional distribution function using the fact that $F_{Y|X=\mathbf{x}}(y) := P(Y \leq y|X = \mathbf{x}) = E[I(Y \leq y)|X = \mathbf{x}]$, where $I$ is the indicator function. The estimated distribution function is thus defined using $\hat{F}_{Y|X=\mathbf{x}}(y) := \sum_{i=1}^n w_i(\mathbf{x})I(y_i \leq y)$. It is this formulation which is also used by the quantile regression forest model, where in that case the weights are determined based on the numbers of trees in the forest which allocate each of the observations to the same leaf node as the query point (Meinshausen and Ridgeway, 2006).

We may then use estimates of the conditional distribution function in order to estimate the corresponding quantiles implicitly based on the relationship $\hat{q}_{\alpha_y(\mathbf{x})}(\mathbf{x}) = y \iff \hat{F}_{Y|X=\mathbf{x}}(y) = \alpha_y(\mathbf{x})$. In our implementation we use a kernel with unbounded

support, which ensures $\hat{F}_{Y|X=\mathbf{x}}$ is strictly increasing, and so one-to-one, and hence this equivalence is valid. To determine an appropriate estimate of $q_\alpha(\mathbf{x})$ for a *chosen* value of $\alpha$, one can simply use interpolation on a grid of pairs of $(y, \alpha_y(\mathbf{x}))$.

## 2.2 A discussion and an instructive example

The approach we propose, although simple, is appealing for its universal applicability in obtaining, among others, prediction intervals for any regression estimator. The univariate kernel smoothing problem can also be performed with a high degree of computational efficiency when compared with the general/multivariate case (Fan and Marron, 1994; Hofmeyr, 2019). It is also worth noting that, although there exist scenarios which are far from pathological in which this approach results in substantial model bias, it also generally enjoys considerably lower estimation variance than the standard non-parametric alternative. One of the primary reasons for this is the simple fact that the subset of the observations described by $\{\mathbf{x}_i|\hat{\mu}(\mathbf{x}_i) \approx \hat{\mu}(\mathbf{x})\}$ is, in general, larger than the subset described by $\{\mathbf{x}_i|\mathbf{x}_i \approx \mathbf{x}\}$, for appropriately defined neighbourhoods of approximation. The subset of observations receiving relatively large weights when relying on our approach is therefore larger than it is in the standard non-parametric approach, leading to a greater effective sample size. The inaccuracy due to bias is therefore mitigated by the lower variance. Furthermore, when the bias is small, i.e., because the approximation $Y|X \overset{D}{\approx} Y|\mu(X)$ holds, at least over a substantial part of the support of $X$, then the lower variance can lead to overall improved accuracy when compared with the alternative. Finally, it is worth noting that a further reduction in variance, when compared with a standard non-parametric estimator, arises from the fact that the non-parametric smoothing is performed in only one dimension. For a multivariate smoother to achieve as low variance as a univariate counterpart, it requires the "neighbourhood of relevance" to be fairly large, which itself may induce estimation bias which even exceeds the model bias of our approach.

**Example:**

Here we illustrate some of the points discussed above graphically, by means of a toy example. At the base we use the many to one function $g(x) = \cos(2.5x) + 0.2x^2$, and define the mean function, $\mu : \mathbb{R}^5 \to \mathbb{R}$ as $\mu(\mathbf{x}) = g(\mathbf{x}'\mathbf{w})$, where $\mathbf{w} \in \mathbb{R}^5$ was determined randomly. A function like $\mu$, which varies only in one direction (i.e., in the direction of $\mathbf{w}$) is referred to as a ridge function. The benefit of using a ridge function as the underlying mean function is that it makes visualising the multivariate function, and estimates related to it, more straightforward, by considering plots of the response against the univariate $\mathbf{x}'\mathbf{w}$.

The main difficulty our approach may face is when there are multiple distinct points, say $\mathbf{x}_1, ..., \mathbf{x}_k$, for which $E[Y|X = \mathbf{x}_i] = E[Y|X = \mathbf{x}_j] \;\forall i, j = 1, ..., k$, but for which the complete distributions of $Y|X = \mathbf{x}_1, ..., Y|X = \mathbf{x}_k$ differ substantially. The equality of the conditional expectation at multiple points is accommodated, in this example, by the design of $\mu$, and we start by considering two scenarios resulting in vastly differing residual distributions, and hence ones in which the assumption on which our approach is based is violated to a very large degree, (i) where the residual distributions have opposite skewness depending on the sign of $\mathbf{x}'\mathbf{w}$, as well as variation increasing in the magnitude of $\mathbf{x}'\mathbf{w}$: $\frac{3\text{sign}(X'\mathbf{w})}{(1+|X'\mathbf{w}|)}\epsilon + 1|X \sim Exp(1)$; and (ii) where the variation of the residual increases substantially with the magnitude of $\mathbf{x}'\mathbf{w}$: $\frac{3}{(0.5+(X'\mathbf{w})^2)}\epsilon + 1|X \sim Exp(1)$.

Figure 1 shows estimated 95% prediction intervals for scenario (i). The plots show the estimated prediction intervals from the quantile regression forest model (QRF), a gradient boosted ensemble of trees (QGB) arising from the minimisation of the sample quantile error (4), and estimates arising from our approach applied to the outputs of a standard regression forest (qRF) and an estimate for $\mu$ coming from a boosted ensemble of trees (qGB). Hyperparameters were selected in the same manner as in the experiments which are presented in the following section. The solid black line shows the mean function, $\mu$, while the dashed lines show the true 0.025 and 0.975 quantiles of the conditional distributions of $Y|X = \mathbf{x}$. The grey points show a typical sample from the underlying distribution, drawn independently from the samples used to perform the estimation. The jagged dark red lines show the average estimates of the 0.025 and 0.975 quantiles of the conditional distributions from 20 replications, while the faint red lines show the results from all 20 runs. For all 20 runs the values for $\mathbf{x}$ in the training (7500 points) and test (2500 points) sets were kept fixed, and the responses were resampled independently each time.

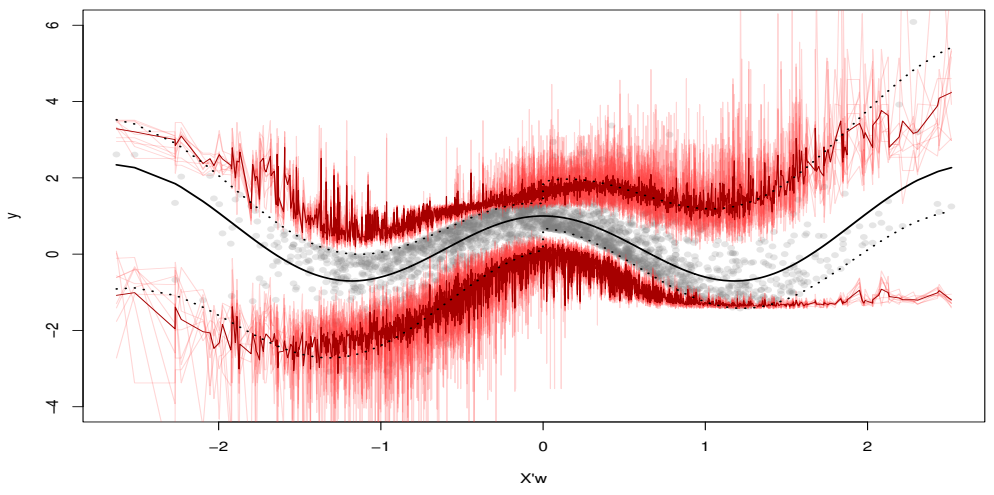

(a) QRF

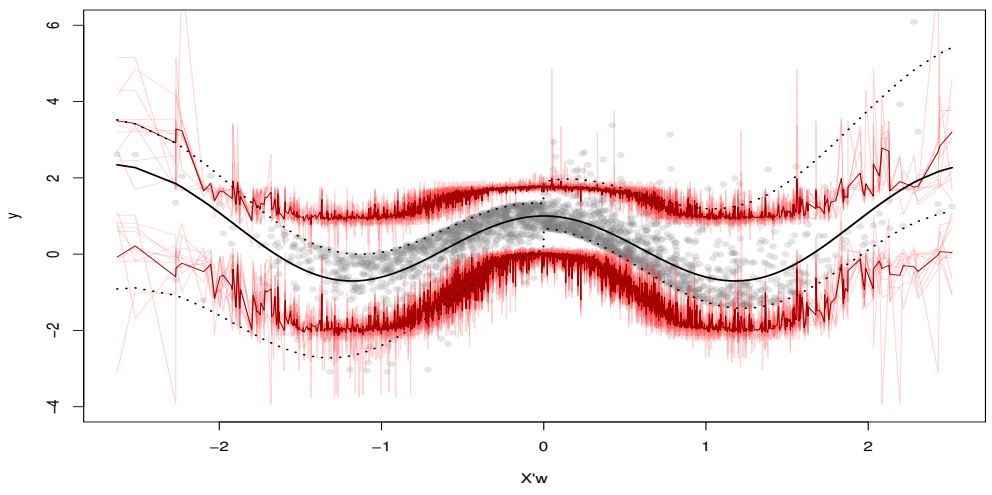

(b) qRF

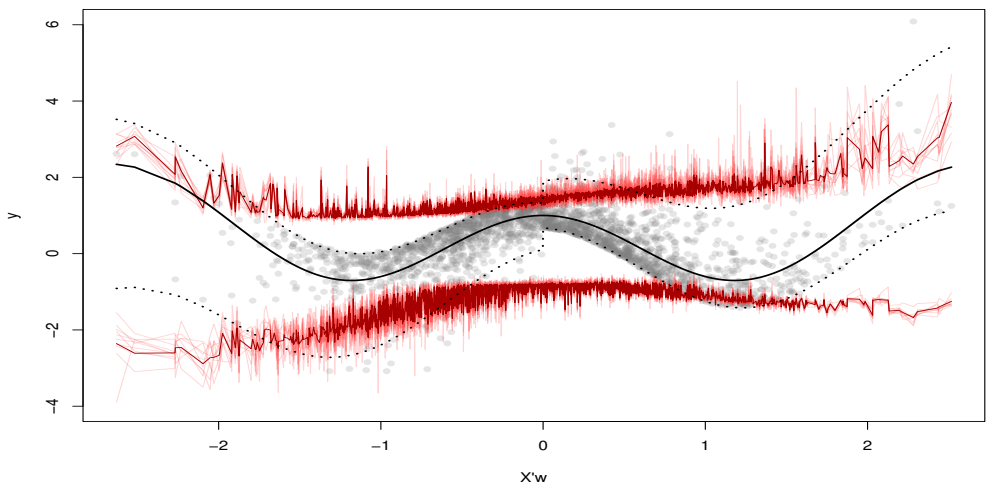

(c) QGB

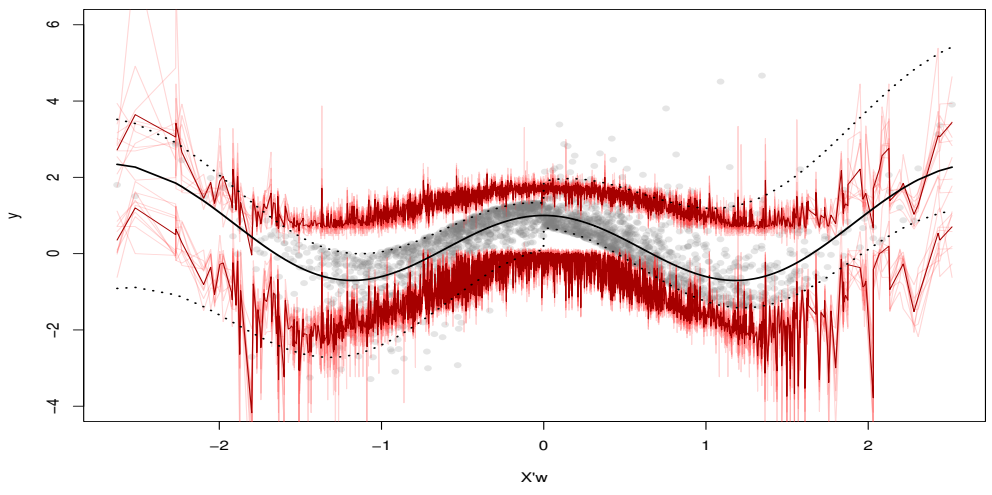

(d) qGB

Figure 1: Estimated 95% prediction intervals under scenario (i). Conditions are designed to illustrate bias of our proposal when regression (mean) function is many-to-one and residual distribution differs greatly (here due to anti-symmetric skewness) at points with same conditional mean.

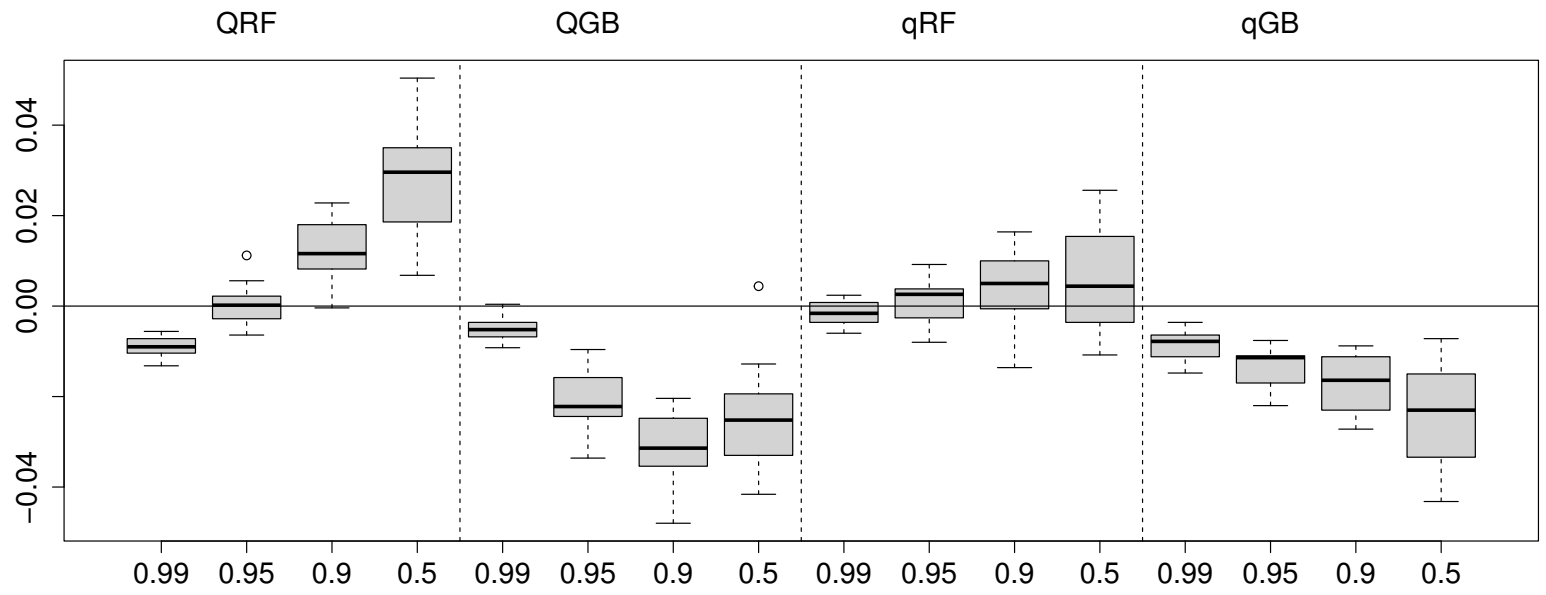

Figure 2: Boxplots of coverage achieved by prediction intervals on test sets from 20 replications under scenario (i). The points contributing to the boxplots are the observed proportions of test cases lying in their prediction intervals, minus the target proportions of $1 - 2\alpha$, for $\alpha \in \{0.005, 0.025, 0.05, 0.25\}$.

As expected, because of the design of the problem, our approach cannot pick up on the anti-symmetry in the skewness of the residual distributions[1]. This is because, in essence, it is merging the conditional distributions at points where the mean values are equal, which are, by design, symmetric about zero. Both QRF and QGB can model this asymmetry, with QRF providing good estimation over much of the domain, but also exhibiting substantial bias in the 0.025 quantile over much of the region $\mathbf{x}'\mathbf{w} \geq 1$. The quantile estimates coming from QRF are derived from (a modified variant of) the standard non-parametric smoothing method, and this bias arises from fairly large neighbourhoods of relevance over that region. The reduction in variance resulting from our simplification is clearly evident in the comparison between QRF and qRF, and despite this example having been designed to illustrate the weakness of our approach the average quantile error achieved on the test sets is only moderately above that of QRF; between 1.00 and 1.15 times greater, where the values for $\alpha$ which were considered are $\{0.005, 0.025, 0.05, 0.25, 0.5, 0.75, 0.95, 0.975, 0.995\}$.

[1]It is very important to note that our approach is not, in general, unable to model asymmetry, but this example was designed in an adversarial manner so that the mean function itself is symmetric, while the residual distribution is perfectly anti-symmetric.

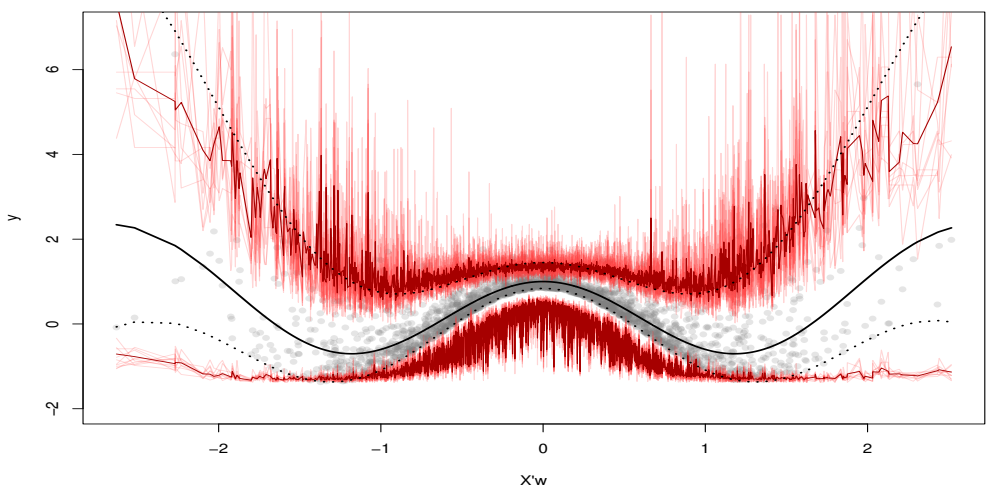

(a) QRF

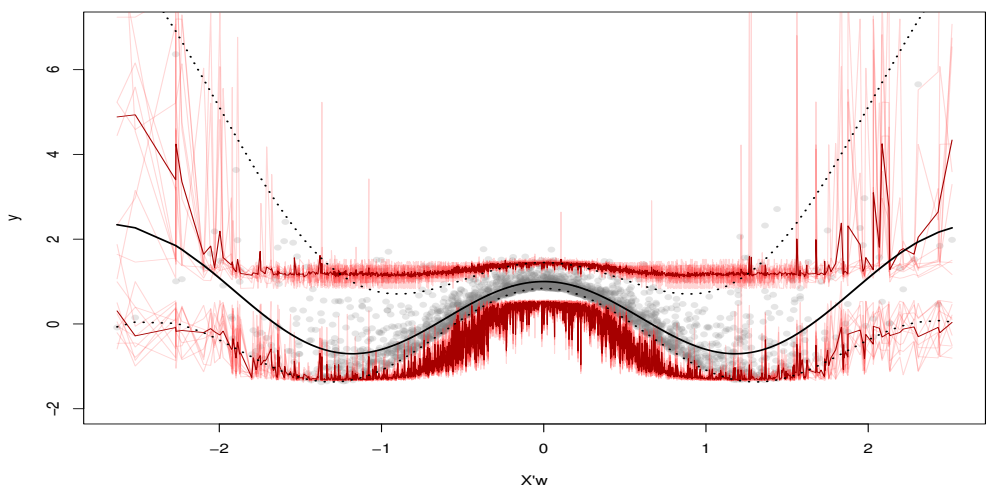

(b) qRF

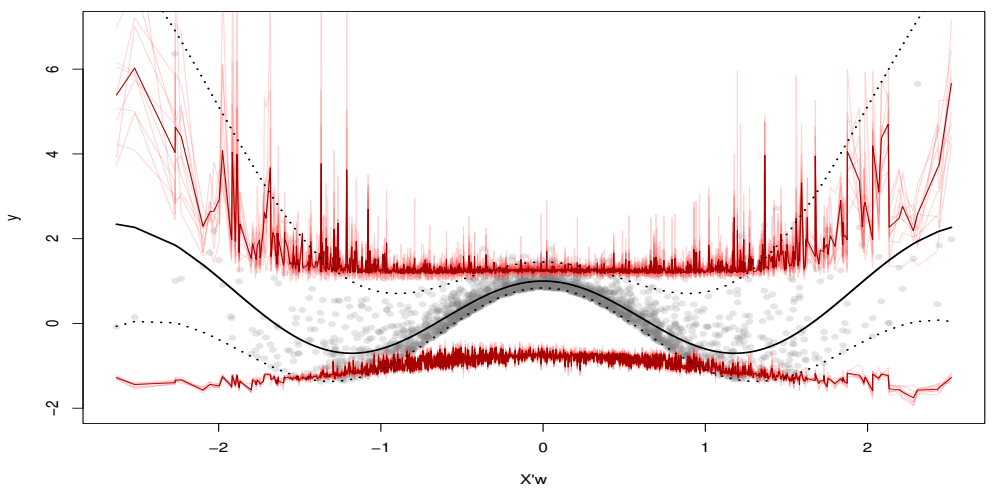

(c) QGB

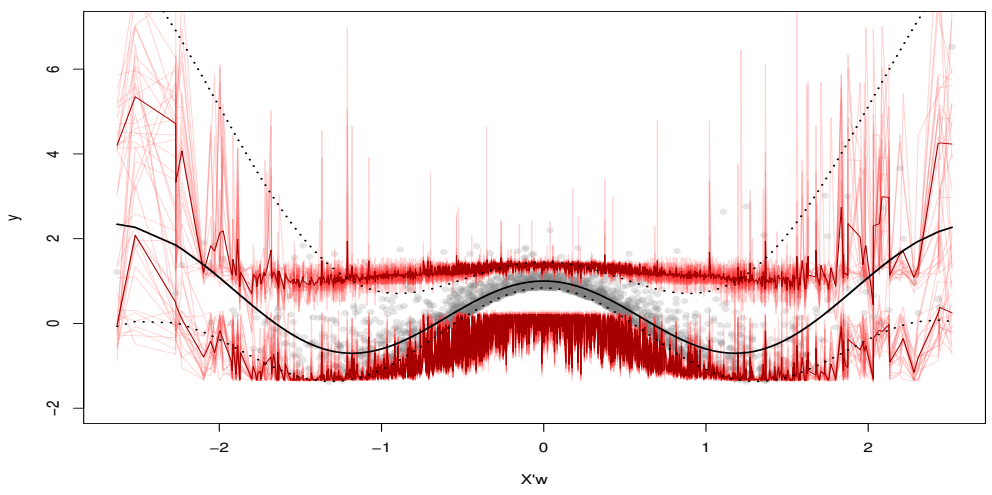

(d) qGB

Figure 3: Estimated 95% prediction intervals under scenario (ii). Conditions are designed to illustrate bias of our proposal when regression (mean) function is many-to-one and residual distribution differs greatly (here due to varying scale) at points with same conditional mean.

The comparison between QGB and qGB does not avail as much intuition, since the QGB estimates do not arise from the standard non-parametric approach. The bias of QGB is here simply a result of the chosen hyper-parameters. In this instance it is qGB which has the greater variance, especially in the 0.025 quantiles. However, as before and despite the adversarial design of the example, the performance is similar to QGB (with relative errors between 0.96 and 1.17). An important point to note is the comparatively higher variance in the estimates coming from qGB over qRF, which is because the smoothing performed in our approach is based on the *estimated* mean function, and hence the variance in *its* estimation is inherited by our estimates.

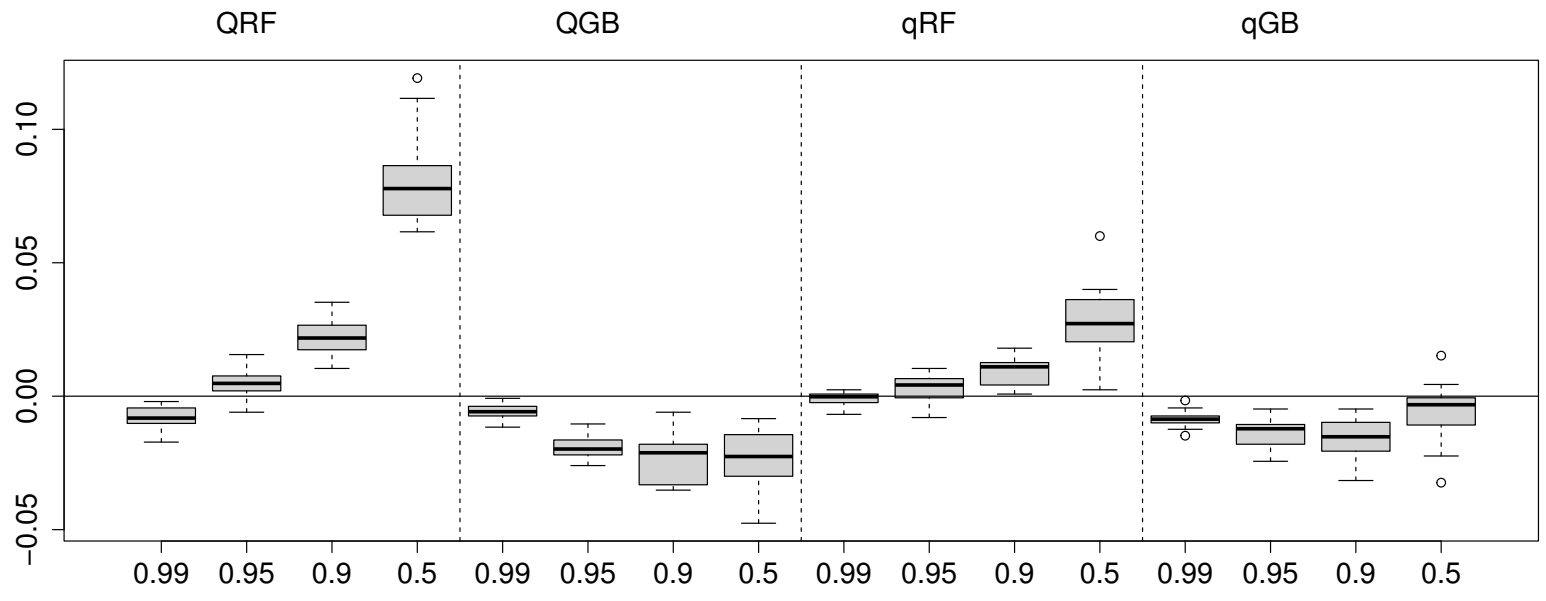

Figure 4: Boxplots of coverage achieved by prediction intervals on test sets from 20 replications under scenario (ii). The points contributing to the boxplots are the observed proportions of test cases lying in their prediction intervals, minus the target proportions of $1 - 2\alpha$, for $\alpha \in \{0.005, 0.025, 0.05, 0.25\}$.

The greater variance in the estimation of $\mu$ coming from the underlying GB models when compared to the RF models is therefore reflected in our estimates.

In addition to the quantile error, it is important to consider the coverage achieved by the estimated prediction intervals on the test sets, since inaccurate coverage shows an inability to adequately capture the uncertainty in the predictions overall. Figure 2 shows boxplots of the error of the coverage achieved by the prediction intervals on the test sets. Each point contributing to a boxplot is the proportion of the test sample lying within the corresponding estimated prediction interval minus the target coverage. Despite the bias of our approach being deliberately accentuated by the design of this example, it still achieves very good overall coverage when compared with the existing approaches.

Figures 3 and 4 show the corresponding plots from scenario (ii). Here it is not the skewness which presents a difficulty for our approach, but rather the scale. That is, there are multiple points, $\mathbf{x}$, for which $Y|X = \mathbf{x}$ has the same mean but vastly different scale. This is reflected in the substantial underestimation of the width of the prediction intervals in the extremes of Figures 3(b) and 3(d). Note that the density of points near the value zero along direction $\mathbf{w}$ is much greater than at larger absolute values, and so the estimated prediction interval width near the extremes is dominated by the small scale from points near zero since the mean values are similar. This is important to note, as when accurate estimation of, especially extremal prediction intervals is important in regions of data sparsity, as may arise in applications involving risk, for example, the proposed approach may not be the most appropriate. Once again, however, despite this example being designed to illustrate a limitation in the proposal, the performance is reasonable, with the test quantile error for qRF being between 0.97 and 1.31 that of QRF, and the error of qGB being between 0.83 and 1.23 that of QGB.

Next we consider cases which are (more) favourable for our approach. In order to mitigate the effects of the many-to-one mean function we add a linear term, so that now $\mu(\mathbf{x}) = g(\mathbf{x}'\mathbf{w}) + 1.25\mathbf{x}'\mathbf{w}$. Scenario (iii) corresponds with the same residual distributions as scenario (i), but with the mean function modified as described above. The estimated 95% prediction intervals can be seen in Figure 5.

The result of the modification is that the mean function is now at most three-to-one within the range of the observations, where previously it was at most four-to-one. More importantly, however, the degrees by which the residual distributions differ at points with equal mean are, in general, lesser than in scenario (i). This is especially the case for values $\mathbf{x}'\mathbf{w} \geq 0$.

In scenarios (i) and (ii) the mean function was symmetric and so applying linear models would not be informative. Now, with the addition of a linear term, we also include quantile estimates from linear models arising from minimising (4), denoted QLM, and from applying our method to the outputs from a linear estimate for the mean, qLM.

The simple modification to the mean function more-or-less reverses the comparative performance in terms of quantile accuracy, with the test quantile errors from qRF now being between 0.88 and 1.01 times those from QRF, and qGB obtaining errors between 0.70 and 0.99 times those of QGB.

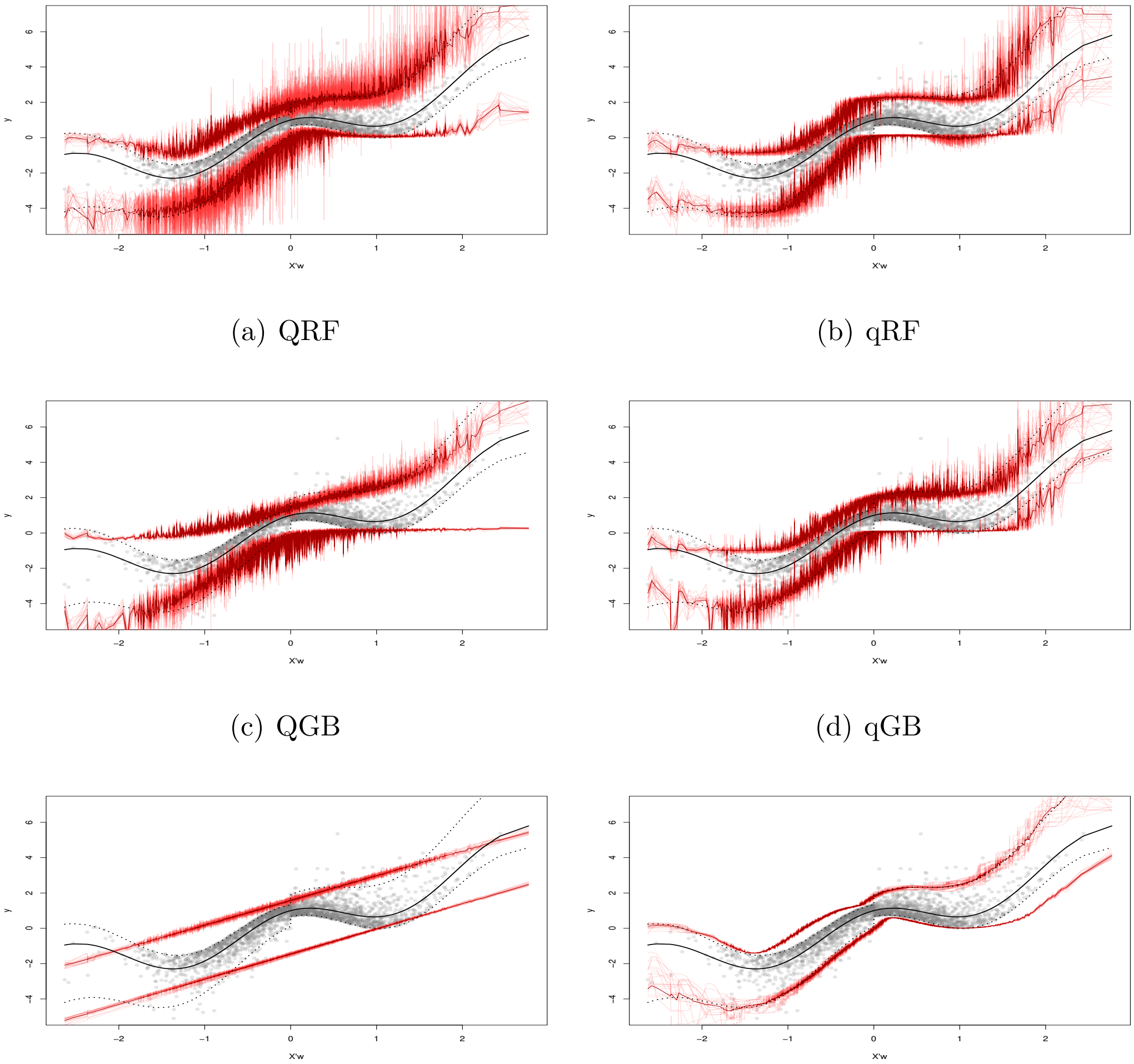

(a) QRF (b) qRF

(c) QGB (d) qGB

(e) QLM (f) qLM

Figure 5: Estimated 95% prediction intervals under scenario (iii). Conditions are designed to illustrate reduction in bias of our proposal (when compared to scenario (i)) when residual distribution differs less for points with the same conditional mean.

The case of qLM is an interesting one, and one which is important to discuss. The accuracy on the 95% prediction interval is clearly very high, and indeed its quantile errors are substantially lower than any of the other methods. This is an ideal scenario in which to apply our approach to the linear model, however, since the non-linearities in the mean function as well as the variations in the residual distribution lie in the same direction as the linear component, and hence equal function values in the estimated linear component will reveal relevant information about the variation in the conditional quantiles.

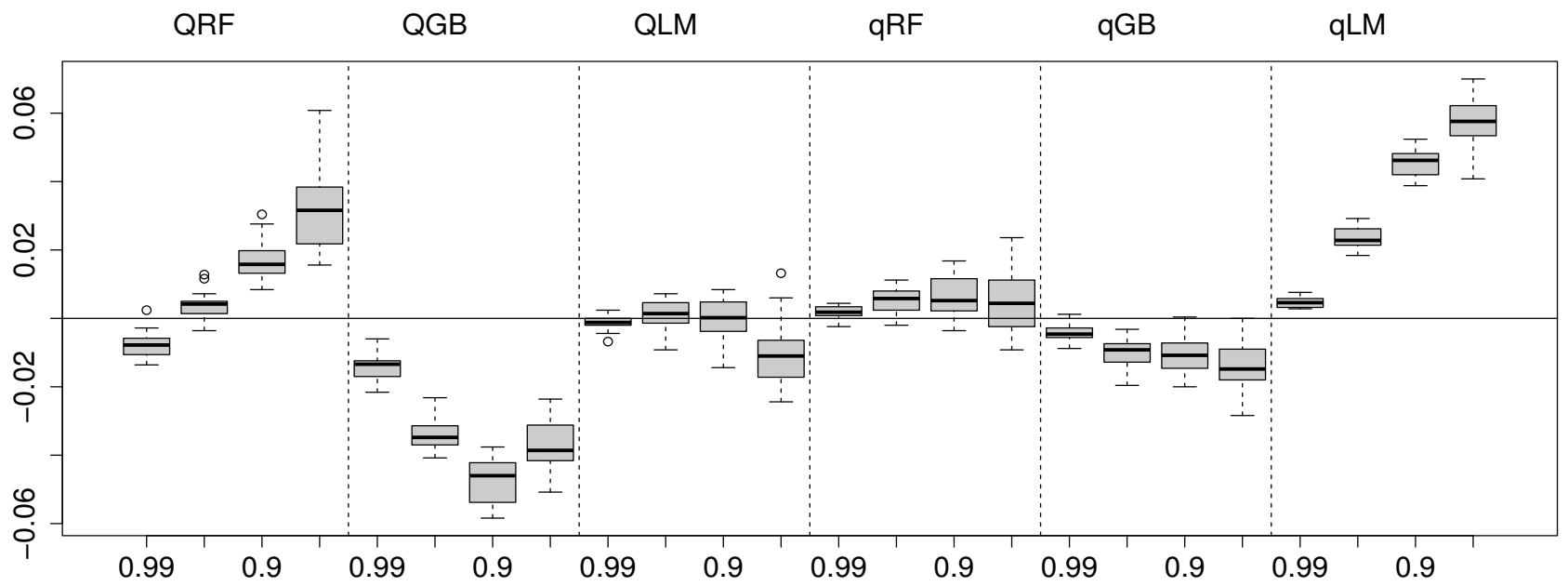

Figure 6: Boxplots of coverage achieved by prediction intervals on test sets from 20 replications under scenario (iii). The points contributing to the boxplots are the observed proportions of test cases lying in their prediction intervals, minus the target proportions of $1 - 2\alpha$, for $\alpha \in \{0.005, 0.025, 0.05, 0.25\}$.

The coverage levels from applying our method to the outputs from the random forest and the boosted model are substantially more accurate than the existing alternatives. The linear model, however, again requires commentary. Although the quantile accuracy is high, the qLM model suffers a mixture of oversmoothing and model bias in the interval roughly between -1 and 0 along the direction $\mathbf{w}$, which is where the data are dense and the true quantile intervals are narrow. It is this oversmoothing, coupled with the high density of observations in this region, which results in considerable over-coverage.

Scenario (iv) corresponds with the same residual distributions as scenario (ii), in

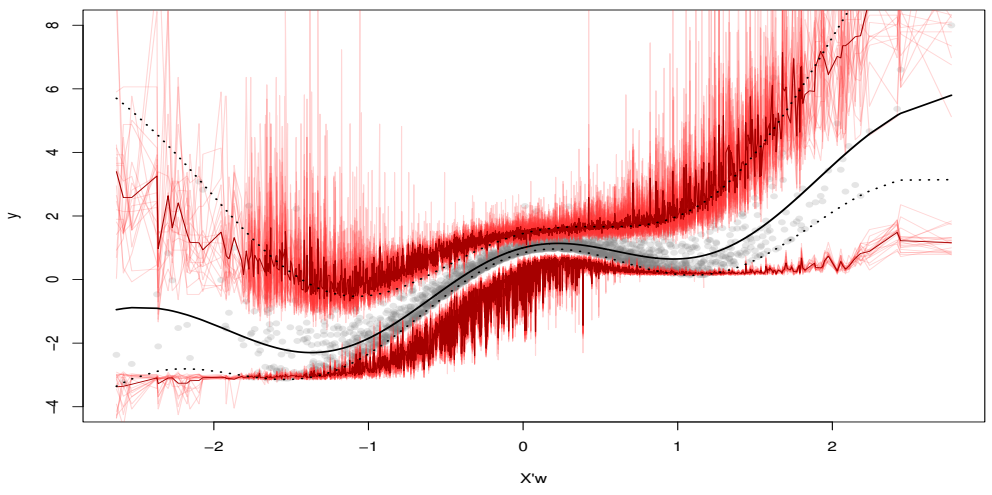

(a) QRF

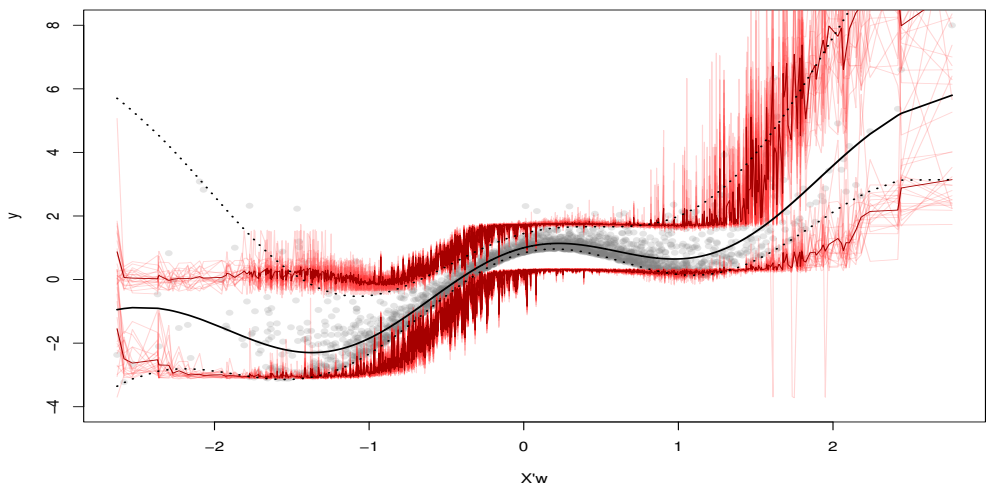

(b) qRF

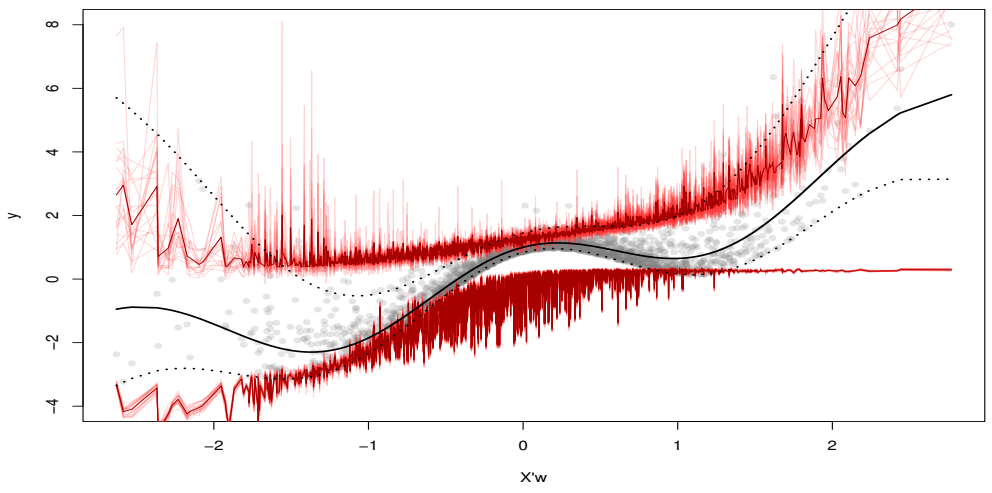

(c) QGB

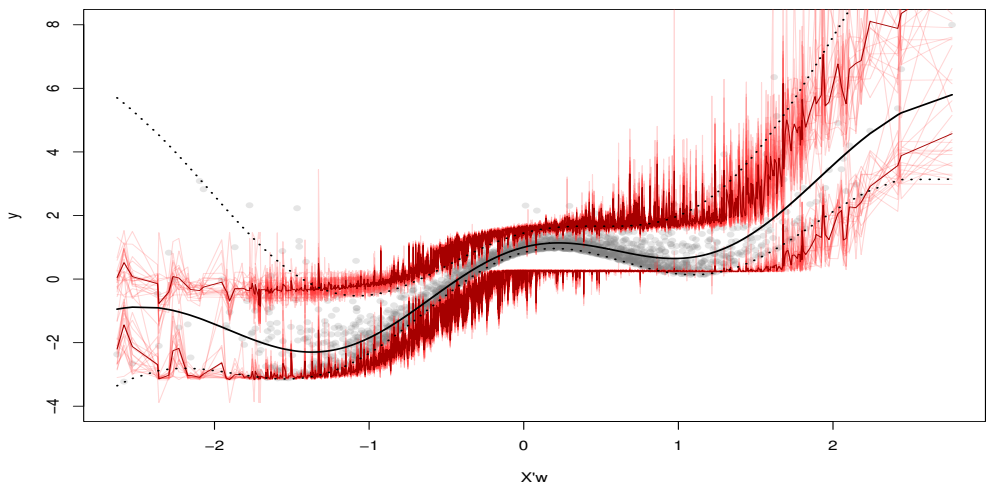

(d) qGB

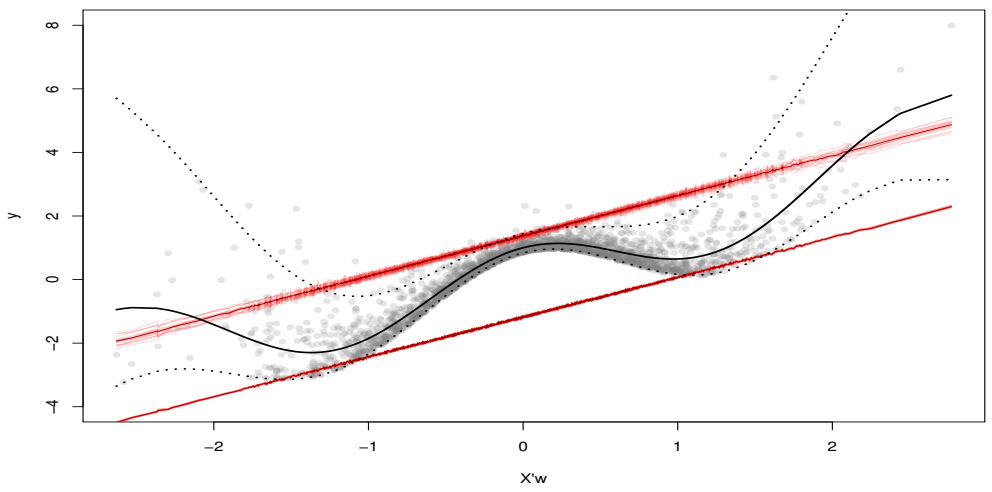

(e) QLM

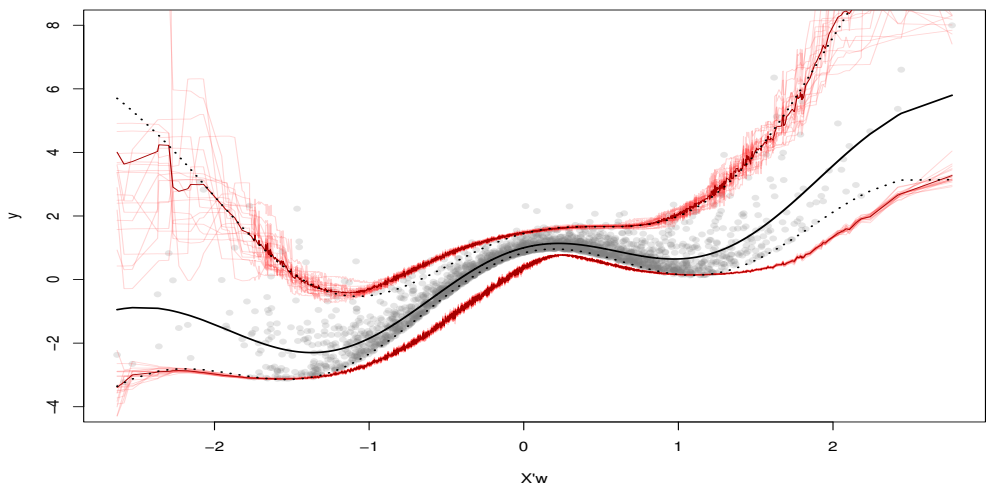

(f) qLM

Figure 7: Estimated 95% prediction intervals under scenario (iv). Conditions are designed to illustrate reduction in bias of our proposal (when compared to scenario (ii)) when residual distribution differs less for points with the same conditional mean.

which the scale increases substantially with the magnitude of $\mathbf{x}'\mathbf{w}$, but with the addition of the linear term in the mean function.

Figures 7 and 8 show respectively the estimated 95% prediction intervals along direction $\mathbf{w}$ and the coverage achieved on the test sets for this scenario. The main region of bias for our approach is where $\mathbf{x}'\mathbf{w} <\approx -1$, where the function is two-to-one but the residual scale differs considerably for some pairs with equal mean. This can be seen in Figures 7(b) and 7(d), where in the extreme left of the plots the width of the intervals is severely under-estimated. Unlike in scenario (ii), however, the accuracy over the rest of the region of observations is fairly high, and qRF has relative errors between 0.95 and 1.13 those of QRF while qGB substantially outperforms QGB in general with relative errors between 0.52 and 1.04. The same comments about the linear model as were made for scenario (iii) are again relevant, with the accuracy being extremely high because of the design of the experiment, but the over-smoothing resulting in over-coverage. As has been the case in general, qRF and qGB achieve reliably accurate coverage in comparison with the existing methods.

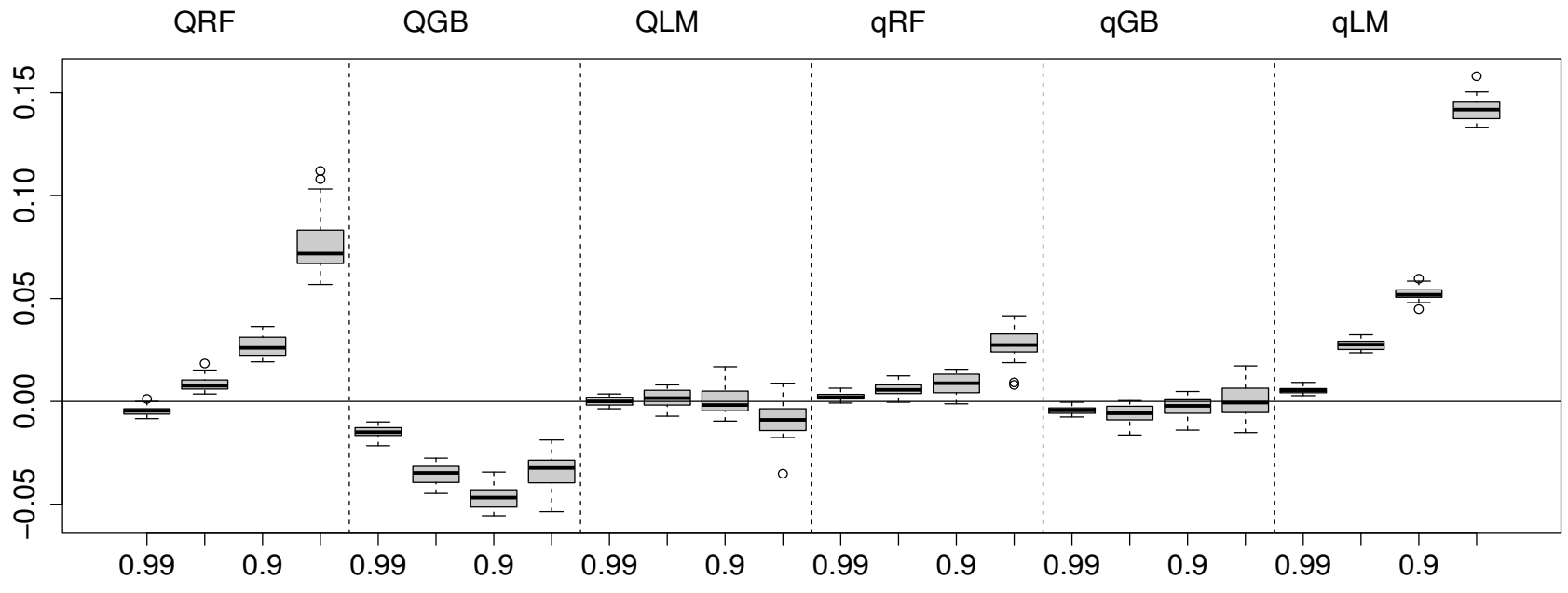

Figure 8: Boxplots of coverage achieved by prediction intervals on test sets from 20 replications under scenario (iv). The points contributing to the boxplots are the observed proportions of test cases lying in their prediction intervals, minus the target proportions of $1 - 2\alpha$, for $\alpha \in \{0.005, 0.025, 0.05, 0.25\}$.

In the final scenario, scenario (v), we consider a case which is strictly favourable for our approach. We return to the same mean function as in scenarios (i) and (ii), but design the residual distribution to be fully determined by the mean, in that $\frac{6\text{sign}(\mu(X))}{1+|\mu(X)|}\epsilon + 1|X \sim Exp(1)$. As a result, not only is estimation in our approach made more efficient by performing the quantile estimation using a univariate smoother (albeit one dependent on a multivariate estimator for the mean), the equality of the residual distribution at points where the mean is equal allows our approach to effectively merge the samples from these conditional distributions to its benefit by further increasing the effective sample size. This can be seen especially well in Figures 9(a) and 9(b), where qRF exhibits substantially lower variance than QRF. It also shows lower bias, again since the smoothing is performed in one dimension rather than, in this example, five. The qGB estimates show much less bias than those from QGB, but again the variation is higher. The boxplots in Figure 10 also once again show the relatively accurate coverage of our approach. The average test quantile errors from qRF are between 0.80 and 0.98 those of QRF, and the qGB errors are between 0.78 and 0.98 those of QGB. Both models derived from the random forest have lower error than both qGB and QGB.

# 3 Experiments on Publicly Available Benchmark Data

As in the previous section, we consider quantile estimates using the proposed approach arising from the outputs from linear regression (LM), boosted ensembles of regression trees (GB) and random forest regression models (RF). We continue with the naming convention from the last section, where QRF; QGB and QLM will be used to refer to the existing approaches and qRF; qGB and qLM refer to our approach applied to the outputs from the three different base models.

Although our approach is applicable to the outputs from any regression model, these three models cover a large part of the spectrum of flexibility, with LM giving rise to very simple models and both RF and GB being (potentially) made very flexible, but with their estimation regularised in very different ways (through bagging and so-called “slow learning”, respectively).

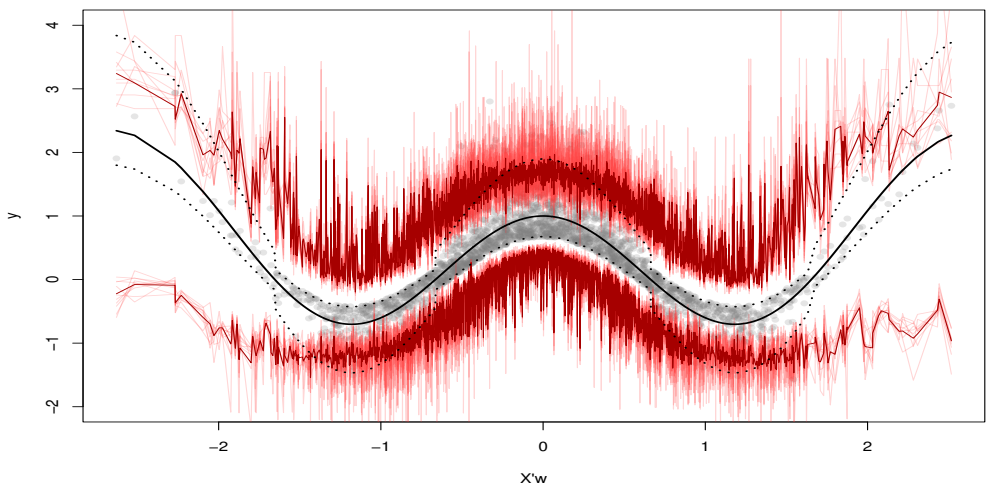

(a) QRF

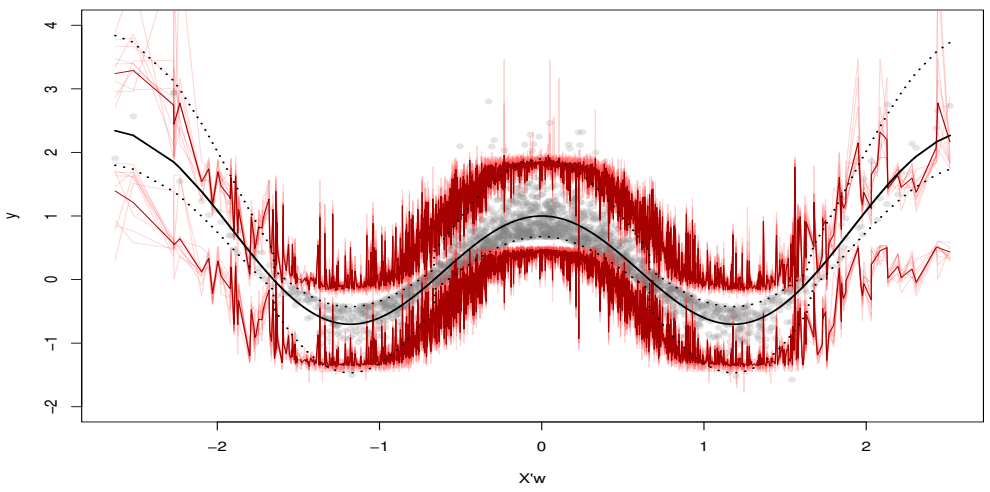

(b) qRF

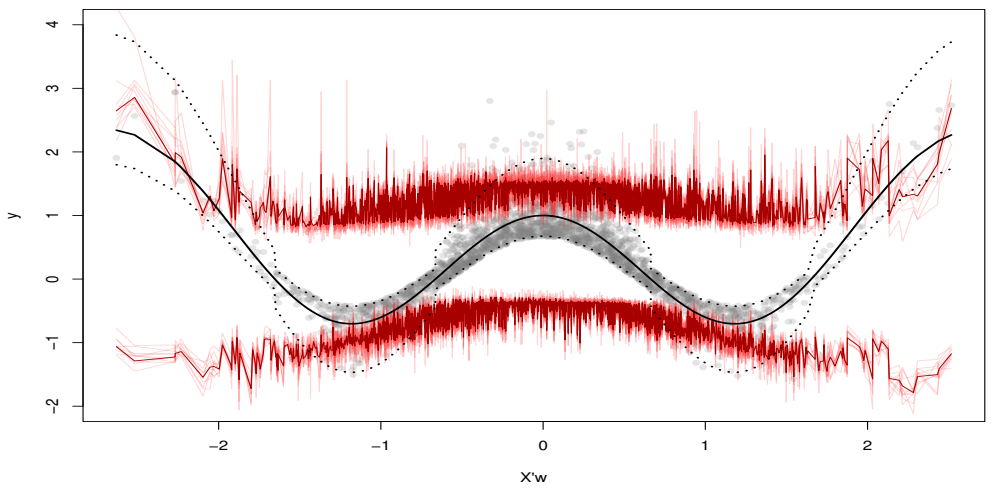

(c) QGB

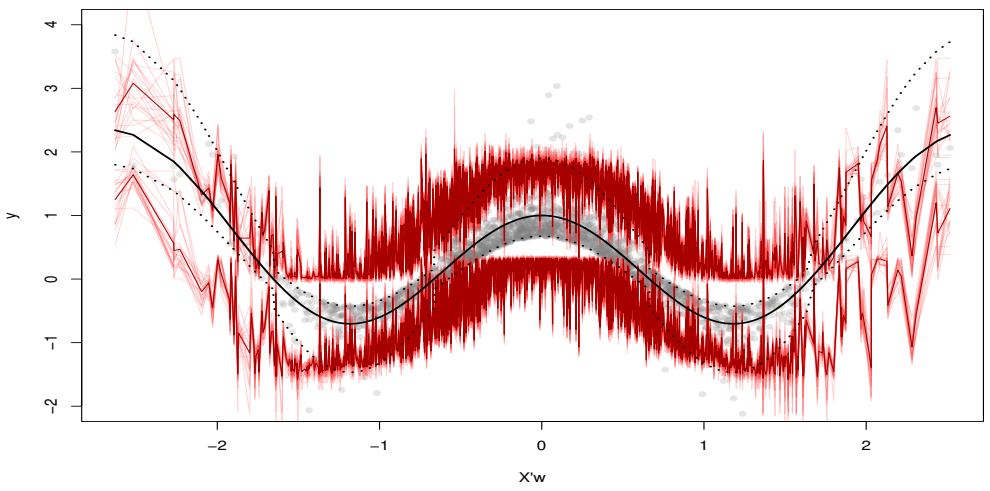

(d) qGB

Figure 9: Estimated 95% prediction intervals under scenario (iii). Conditions are designed to illustrate reduction in variance compared with other non-parametric approaches when regression (mean) function is many-to-one and residual distribution is the same for equal mean.

To fit the random forest models we used the `R` (R Core Team, 2021) package **`ranger`** (Wright and Ziegler, 2017), which includes quantile random forest estimation. To estimate mean and quantile functions using gradient boosting we used the package **`gbm`** (Greenwell et al., 2020). Finally, for the linear models we used our own implementations. To perform the kernel smoothing underlying our proposed approach, we used the `R` package `FKSUM` (Hofmeyr, 2022).

To illustrate the generality of our proposal, we apply these models on a large collection of publicly available benchmark data sets taken from the `R` package `pmlbr` (Le et al., 2020). In the interest of computational brevity we excluded data sets with more than 10000 observations[2]. We also excluded data sets with fewer than 500 observations as the subsampling applied in the implementation of the gradient boosting models rendered errors in some instances due to the selections being too small. Finally, since the implementation of our own proposal is geared towards continuous response variables, we excluded data sets which contained fewer than 25 distinct values for the response[3]. These exclusions left 50 data sets which we have used for comparisons, which are summarised in the remainder.

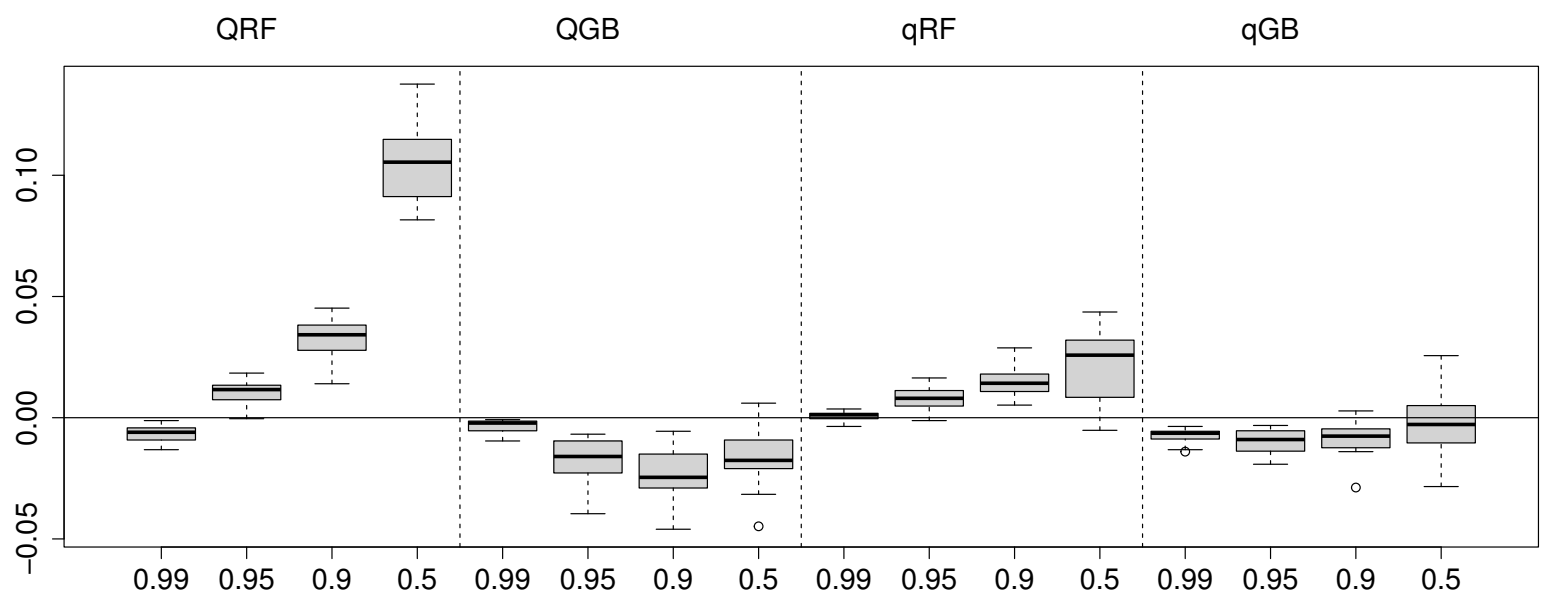

Figure 10: Boxplots of coverage achieved by prediction intervals on test sets from 20 replications under scenario (iii). The points contributing to the boxplots are the observed proportions of test cases lying in their prediction intervals, minus the target proportions of $1 - 2\alpha$, for $\alpha \in \{0.005, 0.025, 0.05, 0.25\}$.

## 3.1 Model Selection

The success of any model is predicated on the assumption that the conditional distribution of $Y$, given $X$, is the same, or at least very similar, within the training data and the unseen "test" cases. Because our proposed estimation method relies on a prior estimation of the mean function, how this translates is that we require that the conditional distribution of $Y$ given $\hat{\mu}(X)$ is not too dissimilar between training and test scenarios. Any substantial overfitting to the training data represents a failure of this requirement. This is particularly problematic in the case of *benign overfitting*, where validation/test error may be low despite overfitting (Bartlett et al., 2020). When utilising our proposed method, it is therefore important to take extra care against overfitting, when performing model selection based on an estimate of prediction accuracy/error. For example, when using (cross-)validation based estimates for prediction performance, it is important to also consider the comparison between training and validation error and to ensure that these are not too different. For the same reason, when utilising a bagged model, such as a random forest, for estimation of the mean function, it is important to use the out-of-bag estimates for the mean evaluated at the training points, since the entire ensemble will tend to overfit on the training cases.

[2]The proposed method does not represent a computational barrier and rather it is the fitting of multiple random forest models which is the most time consuming part of our experiments.

[3]The proposed approach can be applied to discrete responses provided the estimates of the conditional distribution functions are appropriately truncated.

In our experiments we used five-fold cross-validation in order to estimate prediction error for linear models and boosted ensembles. For estimation of the mean function we then selected the models which gave the lowest estimate for prediction error from among those whose training error was no less than 50% that of their validation error. When selecting models for direct quantile estimation on the other hand, i.e., those based on minimising (4), we selected those which minimised the validation error, as is standard. For the random forest models, we simply selected the models which gave the lowest out-of-bag estimates for prediction error.[4]

For the linear models we varied the flexibility by means of a ridge penalty, and considered values of the penalty coefficient between 0 and the largest eigenvalue of the sample covariance matrix of the covariates. The smallest value for the penalty therefore corresponds with the ordinary linear model, while the largest deflates the contribution of the first principal component in the covariates by half (and other components by a greater degree). For the boosted models we only selected the number of terms in the additive models (from between 1 and 1000). The settings for the learning rate (also referred to as shrinkage) and tree depth were kept fixed at 0.02 and 6, respectively. Depth six trees have been reported as an appropriate general purpose setting (Hastie et al., 2009). For the random forest models we only varied the parameter controlling the number of randomly selected covariates as options at each split in one of the trees, sometimes referred to as "mtry". The values considered were $1/3; 1/2; 1; 2$ and $\sqrt{p}$ times the default of $\sqrt{p}$.[5] Other hyperparameters were given their default settings.

[4]In our experience the random forest model is not susceptible to benign overfitting.

For our approach we simply set the bandwidth in the kernel smoothing according to Silverman's rule-of-thumb (Silverman, 1986) applied on the fitted values from the training set. Although this rule of thumb is for the problem of kernel density estimation, the equivalence between the Nadaraya-Watson smoother and the conditional expectation arising from a multivariate kernel density estimate from the response and its covariate(s) (Watson, 1964) justifies its use in this context. Note that, although it is possible to include the subsequent quantile estimation arising from our approach when selecting the model for the mean function, the intention is that the proposed method can be directly applied to the outputs from a well selected regression model. That said, it is worth noting that increasing the bandwidth will tend to increase the width of prediction intervals, and hence increase their coverage, should erring on the side of over-coverage be preferable.

## 3.2 Quantile Accuracy

For the remainder we will use $\mathbf{y}^{te}, \hat{\mathbf{q}}^{te,\alpha} \in \mathbb{R}^{n_{te}}$ to be the vectors of responses and the estimated $\alpha$-quantiles for the test cases. Quantile accuracy performance is assessed based on the test error with respect to the objective in (4), i.e.,

$$\mathcal{E}_\alpha := \frac{1}{n_{te}}\left((1-\alpha)\sum_{i:y_i^{te}<\hat{q}_i^{te,\alpha}}(\hat{q}_i^{te,\alpha} - y_i^{te}) + \alpha \sum_{i:y_i^{te}\geq\hat{q}_i^{te,\alpha}}(y_i^{te} - \hat{q}_i^{te,\alpha})\right).$$

To summarise the results from all 50 data sets used, we first present box-plots, in Figure 11, of the performance, where each point contributing to one of the box-plots is the average performance from 10 random splits of a data set into 70% training+validation and 30% test data. To make the test errors comparable across different data sets we scaled test quantile errors by a factor $\frac{1}{\hat{\sigma}_\epsilon}$, where $\hat{\sigma}_\epsilon$ is an estimate for the residual standard deviation determined based on the smallest root mean squared test error from the three base regression models (RF, LM and GB). For brevity we have combined the results from both tails; taking the averages of $\mathcal{E}_\alpha$ and $\mathcal{E}_{1-\alpha}$.

[5]We take the floor of these values when they are non-integer.

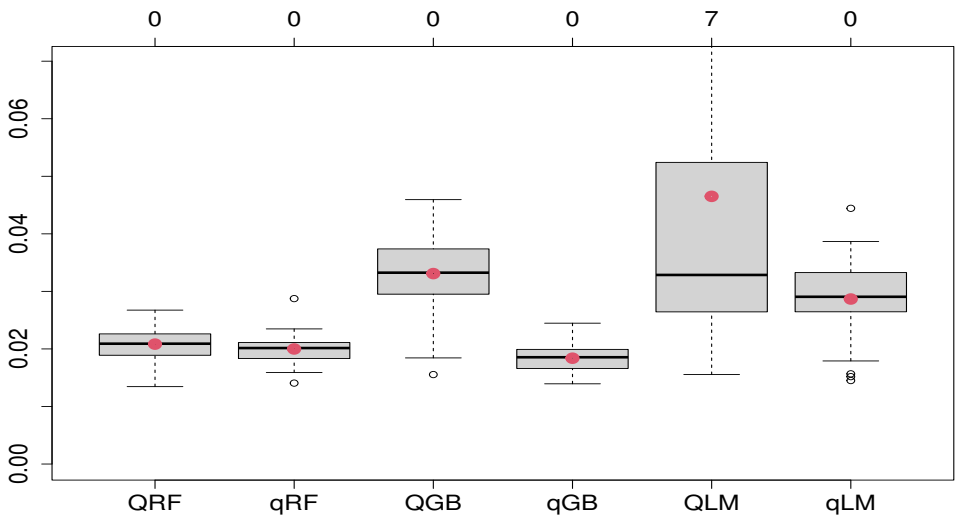

(a) $\alpha = 0.005$

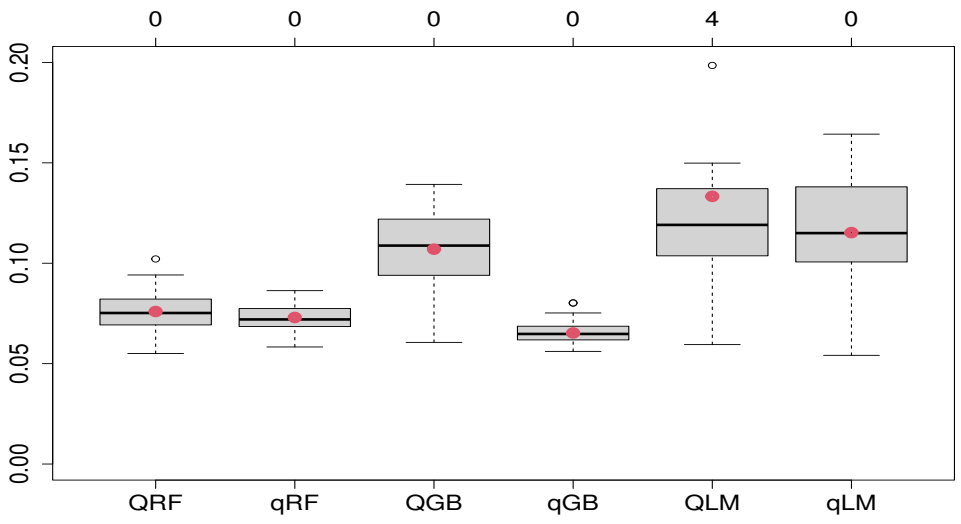

(b) $\alpha = 0.025$

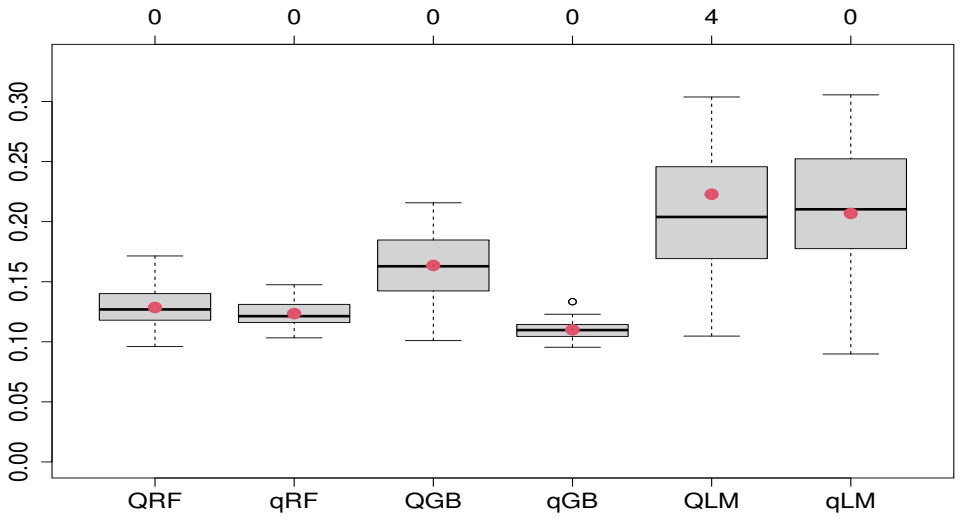

(c) $\alpha = 0.05$

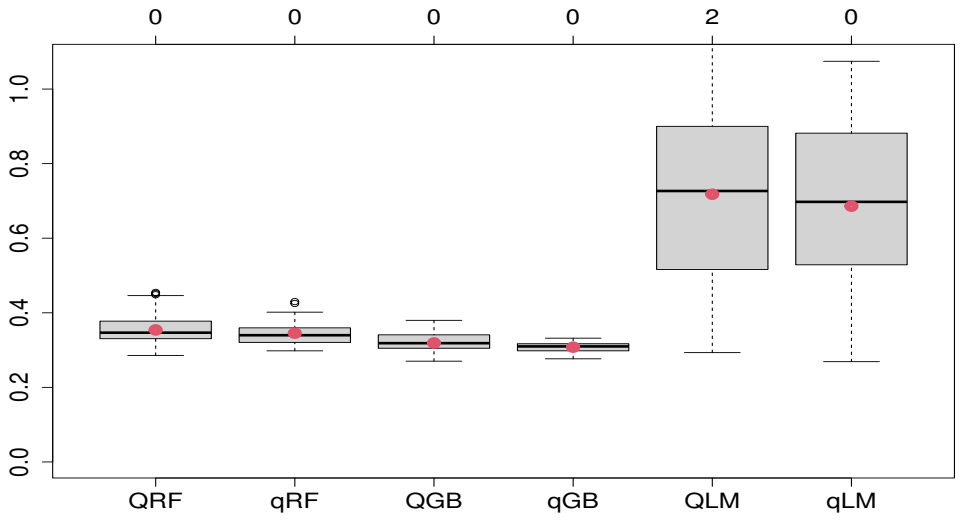

(d) $\alpha = 0.25$

Figure 11: Average of test quantile error from both tails ($0.5\mathcal{E}_\alpha + 0.5\mathcal{E}_{1-\alpha}$) for different values of $\alpha$. Figures along the top axis indicate the numbers of points cropped out of the image. Red dots indicate averages.

The first take-away is that the proposed approach is at least competitive with the existing alternatives for all values of $\alpha$. Furthermore, in most instances the proposed method substantially outperformed the existing alternatives.

Although the boxplots show considerable evidence of the accuracy provided by the proposed method, they do not avail a direct comparison on any single data set.

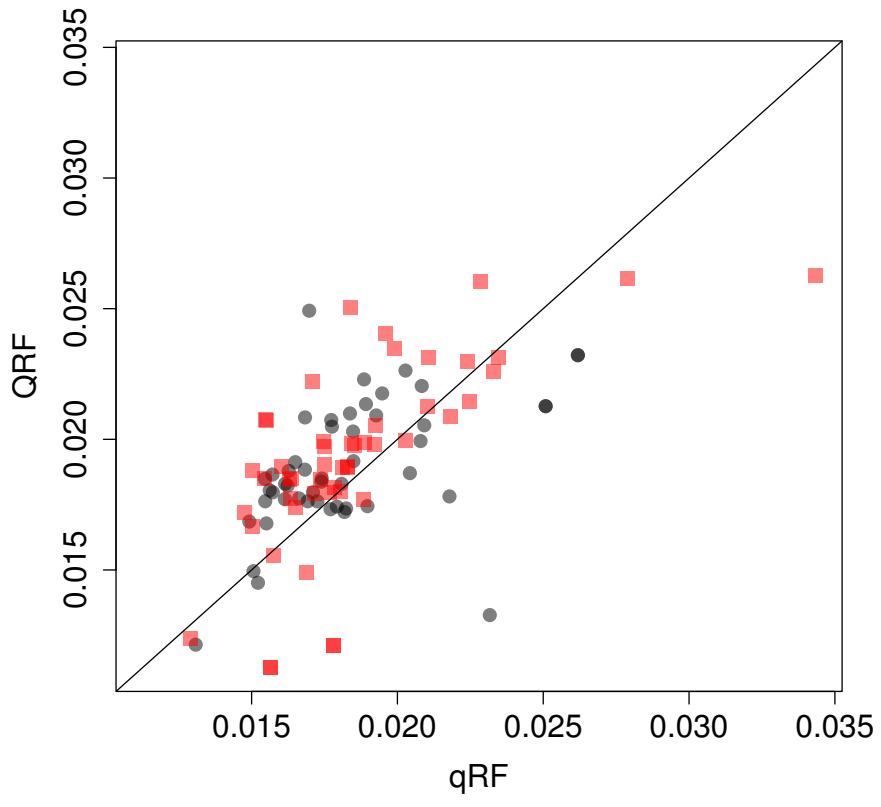

(a) $\alpha = 0.005$

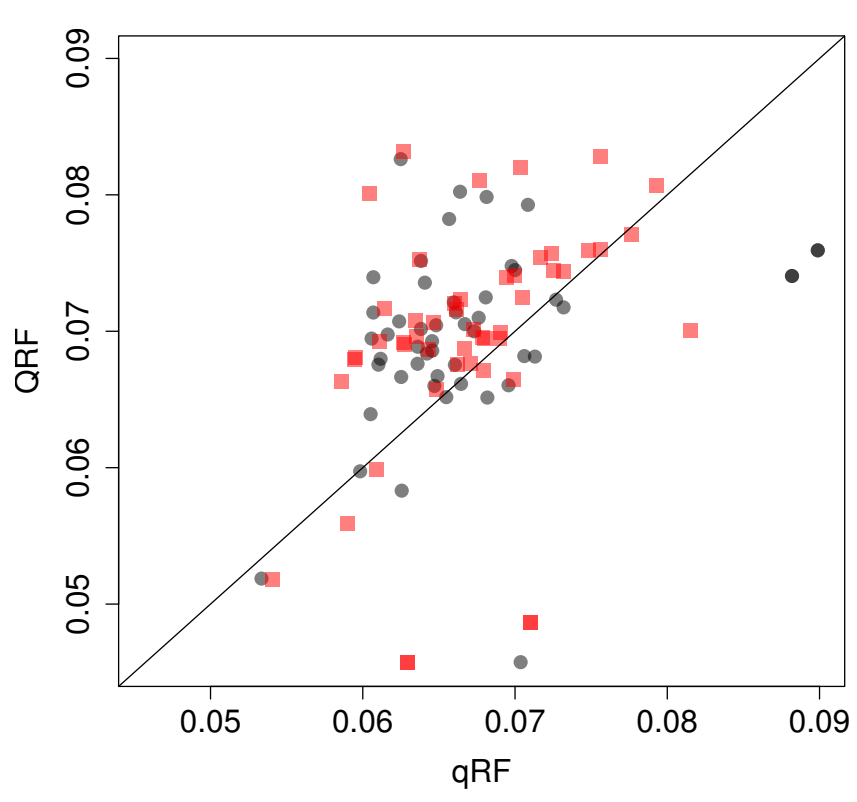

(b) $\alpha = 0.025$

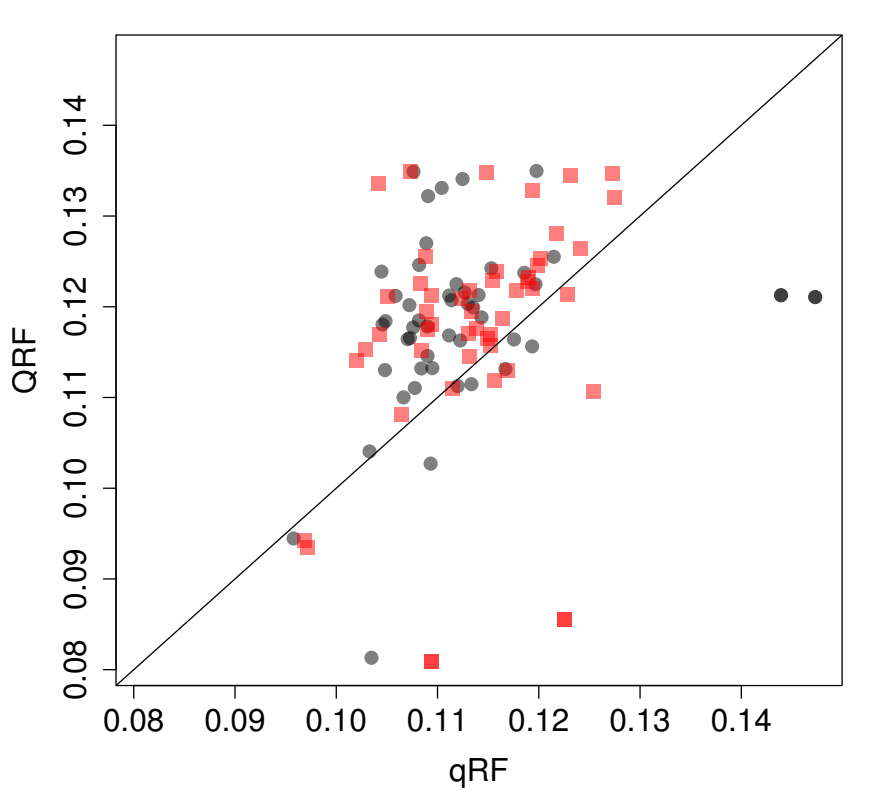

(c) $\alpha = 0.05$

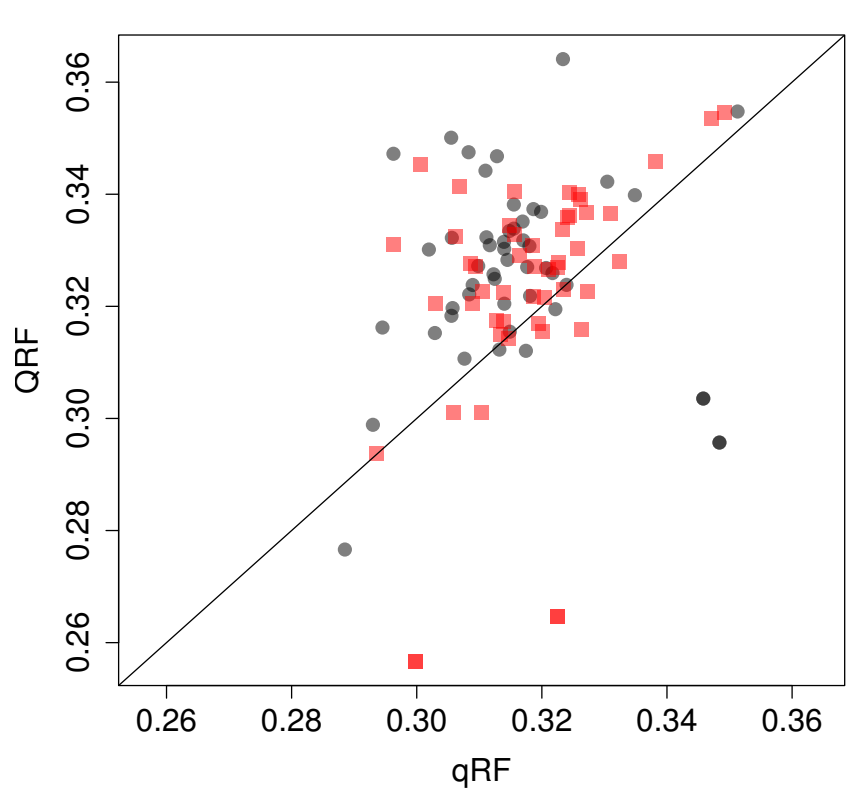

(d) $\alpha = 0.25$

Figure 12: Pairwise comparisons of average test quantile errors obtained by applying our approach to the outputs of RF with those obtained from QRF. ● and ■ indicate left and right tail error respectively. Straight lines indicate equality, with points above the lines indicating higher error for the existing method and points below the lines higher error for our approach.

Figure 12 shows a pairwise comparison of the performance of qRF and QRF on all 50 data sets. In this case we have separated the performance from the left and right tails, and distinguished them based on colour and point character. Points above the diagonals indicate instances where the error for the models shown on the vertical axes (the existing approaches) are greater than those on the horizontal axes (our approach). Not only does the proposed approach provide a high degree of accuracy in general, it also outperforms the existing approach in the vast majority of individual cases. In particular the error from qRF is lower than that of QRF in 74%, 78%, 82% and 82% for $\alpha = 0.005, 0.025, 0.05$ and 0.25, respectively.

For the boosted models, the pairwise comparisons are shown in Figure 13. Here qGB achieves lower error than QGB in 98% of cases for $\alpha = 0.005$ and 0.025 and 94% of cases for $\alpha = 0.05$, and 64% for $\alpha = 0.25$. Finally, the comparisons for the linear models are shown in Figure 14. The comparison here is interesting as the standard approach outperforms the proposed method in the majority of cases; with lower error in 36%, 58%, 66% and 56% of cases for the different values of $\alpha$. However, there are no instances where QLM has substantialaly lower error than qLM, but when qLM achieves lower error it frequently does so by a considerable degree.

## 3.3 Coverage

Once again, accuracy of the quantile estimates does not directly ensure that the prediction intervals appropriately capture the prediction uncertainty from an underlying model. We therefore also consider the empirical coverage of the prediction intervals on the test cases, i.e.,

$$\frac{1}{n_{te}} \sum_{i=1}^{n} I(\hat{q}_i^{te,\alpha} \leq y_i^{te} \leq \hat{q}_i^{te,1-\alpha}).$$

Figure 15 shows boxplots of the average coverage levels achieved by each method on all 50 datasets, where averages are taken over the different training/test splits.

As we found in the simulated example from Section 2, the proposed approach leads to more accurate coverage then the existing approaches. It is also worth noting that, where the prediction intervals arising from minimising (4) tend to under-cover the proposed method shows a tendency to over-cover. Over coverage is generally preferable to under-coverage as it corresponds with a more conservative representation of prediction uncertainty.

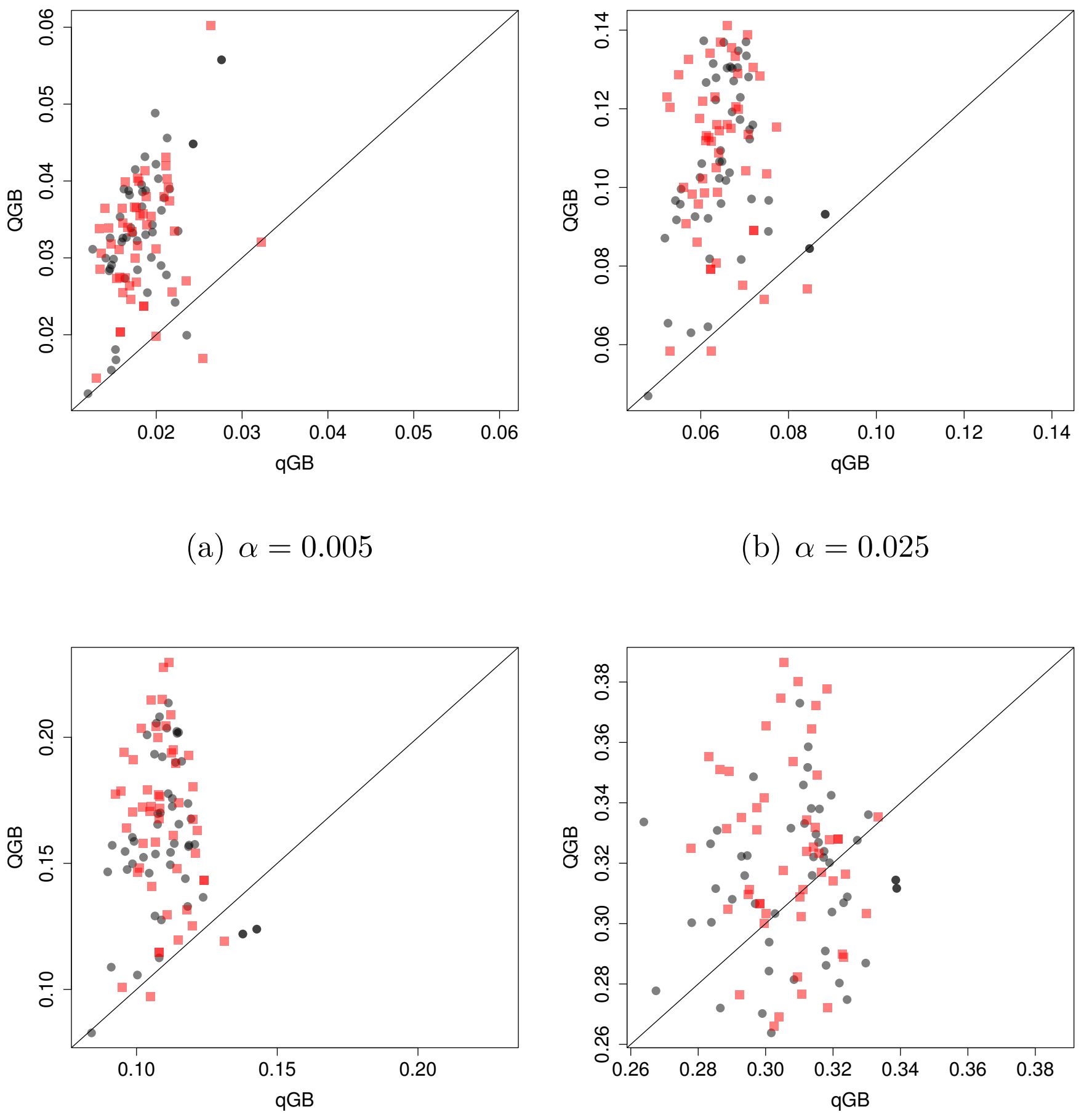

(a) $\alpha = 0.005$ (b) $\alpha = 0.025$

(c) $\alpha = 0.05$ (d) $\alpha = 0.25$

Figure 13: Pairwise comparisons of average test quantile errors obtained by applying our approach to the outputs of GB with those obtained by applying GB to (4). ● and ■ indicate left and right tail error respectively. Straight lines indicate equality, with points above the lines indicating higher error for the existing method and points below the lines higher error for our approach.

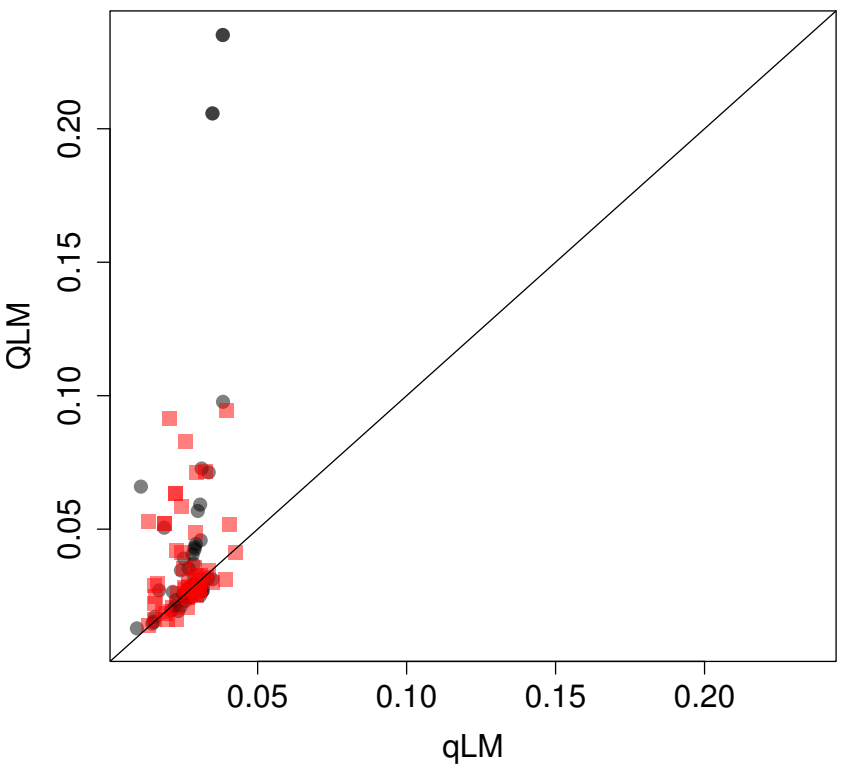

(a) $\alpha = 0.005$

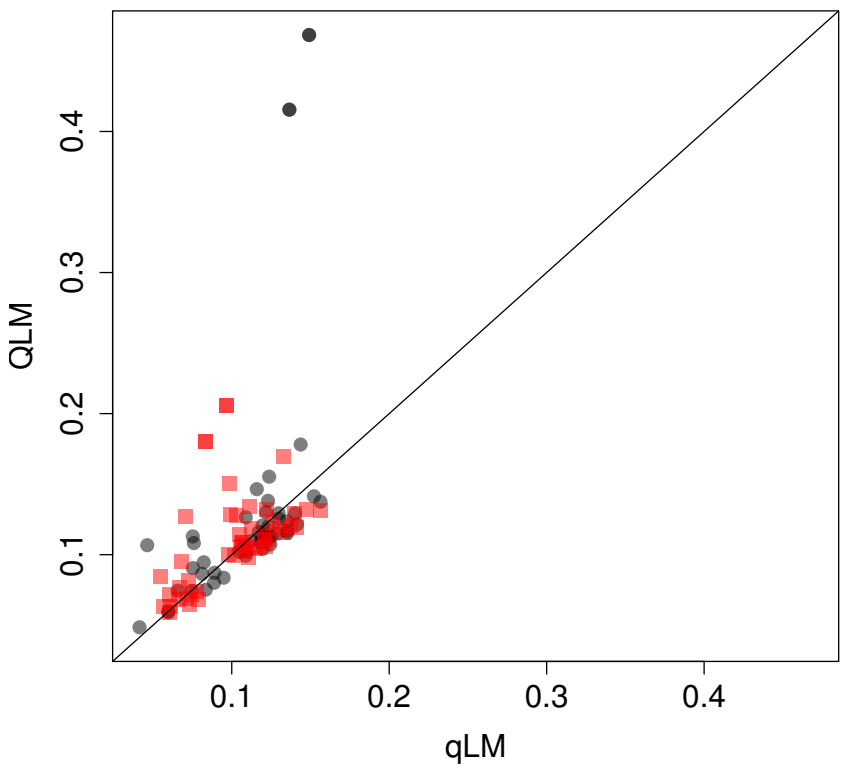

(b) $\alpha = 0.025$

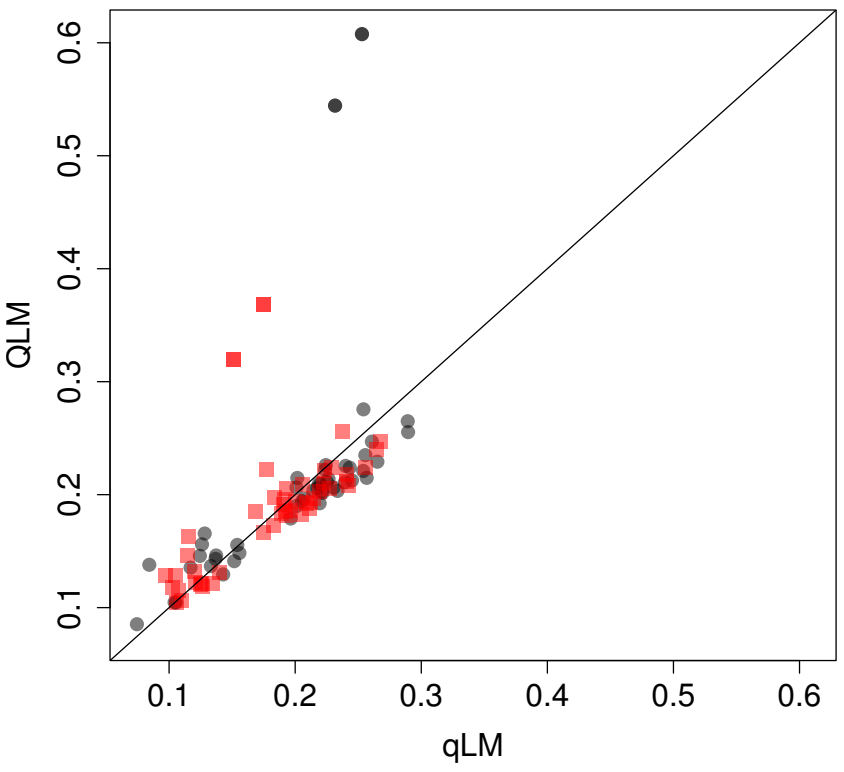

(c) $\alpha = 0.05$

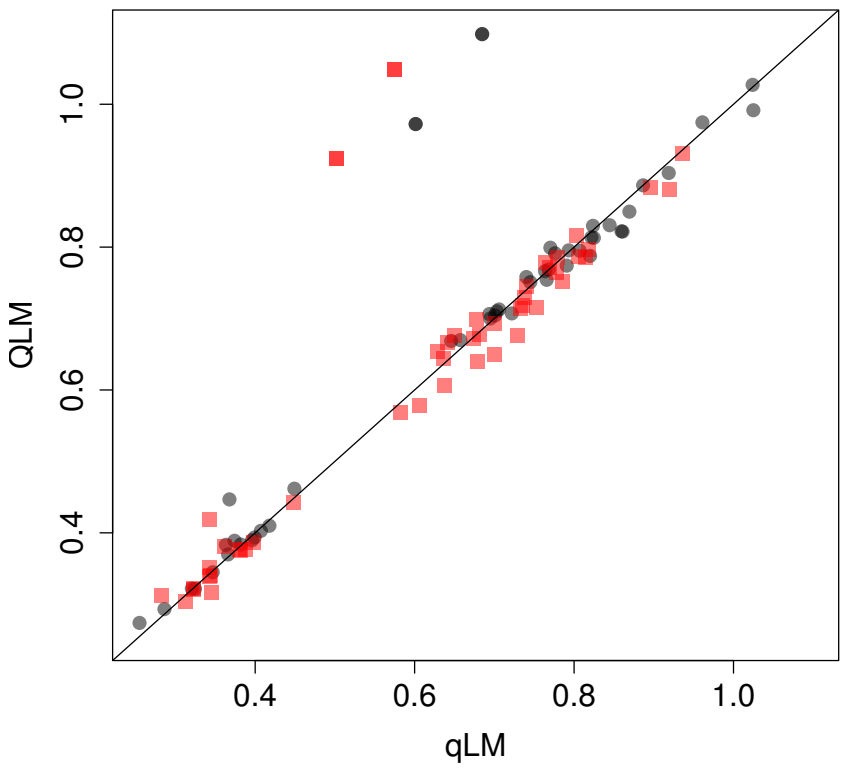

(d) $\alpha = 0.25$

Figure 14: Pairwise comparisons of average test quantile errors obtained by applying our approach to the outputs of LM with those obtained with a linear model minimising (4). ● and ■ indicate left and right tail error respectively. Straight lines indicate equality, with points above the lines indicating higher error for the existing method and points below the lines higher error for our approach.

# 4 Discussion and recommendations

In this work we introduced a simple approach for obtaining detailed information about the conditional distribution of a real-valued response variable, given values for its co-variates. We demonstrated the usefulness of this approach by means of conditional quantile estimation in applications on a large collection of publicly available benchmark data sets. Performance in terms of quantile accuracy and coverage indicate great promise in this approach for leveraging, at least, reliable prediction intervals for quantifying the prediction uncertainty of a given regression model.

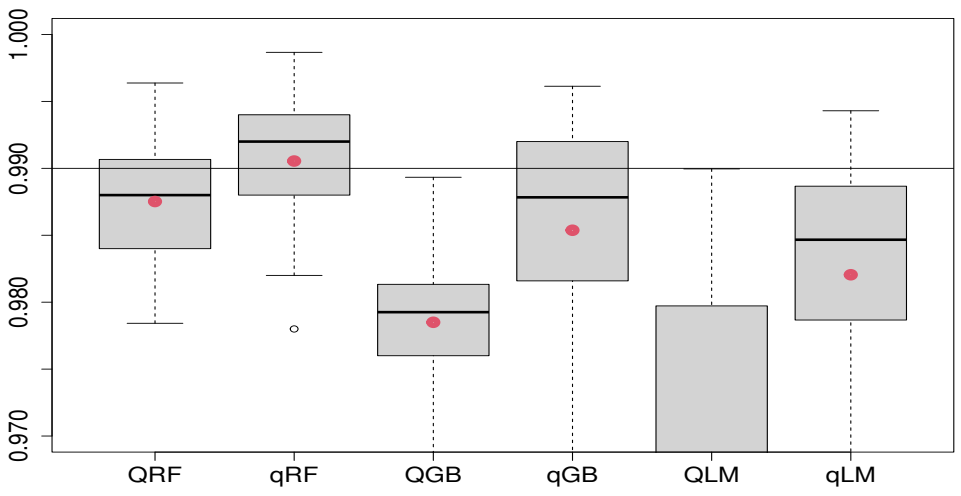

(a) $\alpha = 0.005$

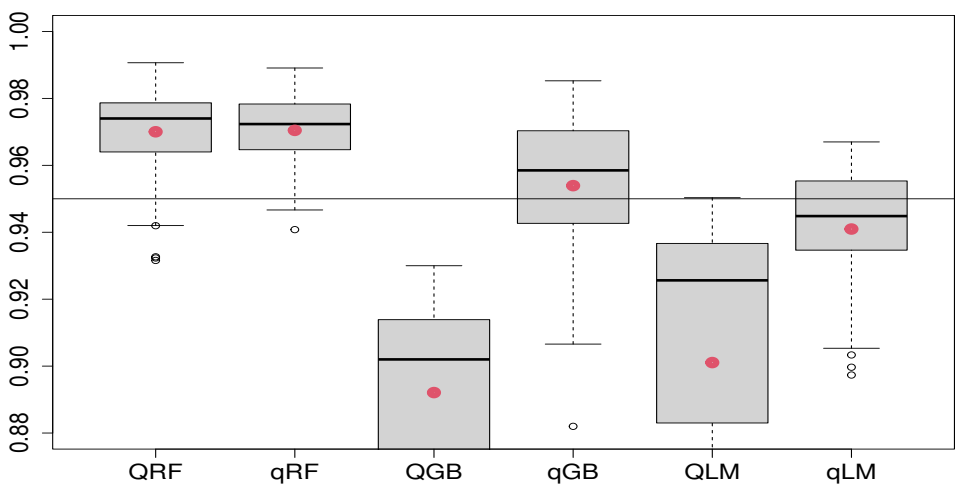

(b) $\alpha = 0.025$

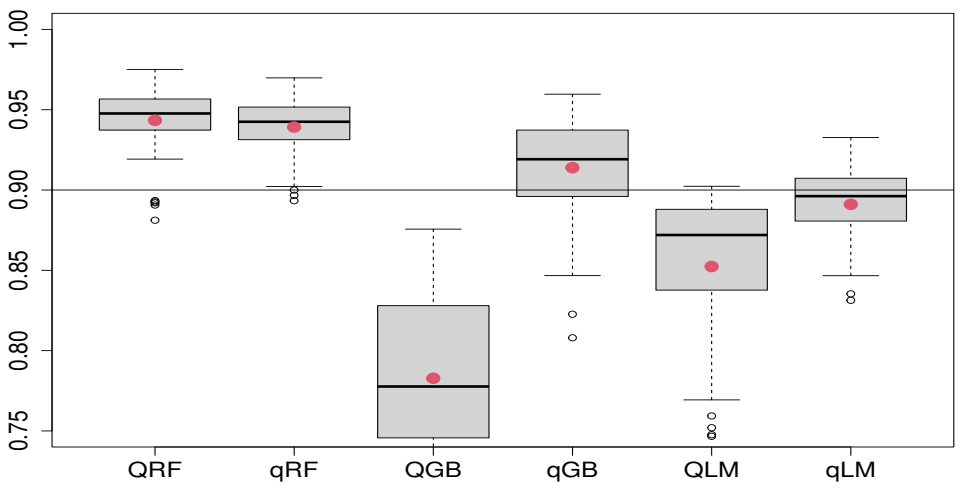

(c) $\alpha = 0.05$

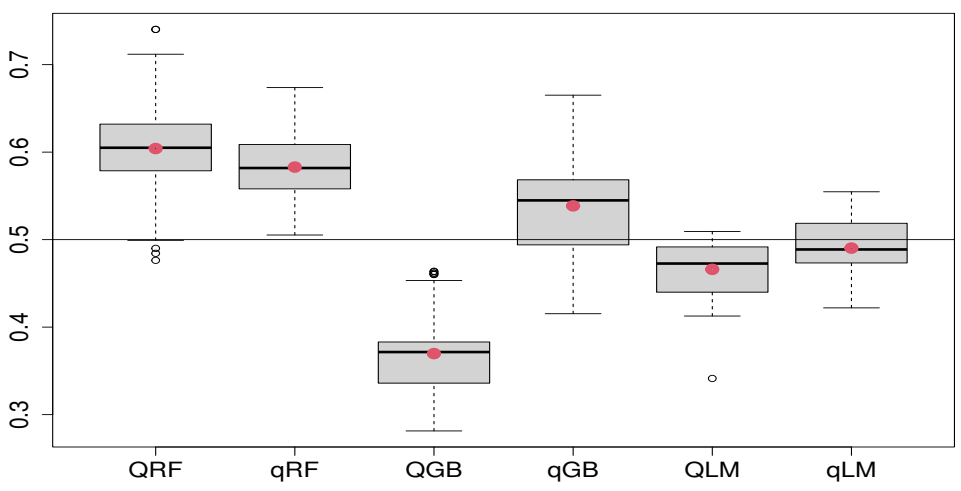

(d) $\alpha = 0.25$

Figure 15: Box-plots of test coverage achieved by estimated prediction intervals of size $1 - 2\alpha$. The horizontal lines show the target coverage level, while the red dots show the averages from each method.

An important caveat relates to the potential problem of benign overfitting when using very flexible regression estimators for the underlying model. Care should be taken when utilising the proposed approach, and it is not recommended when the training and validation errors of a model differ substantially. In our experiments, we used a simple threshold approach, and discarded models whose training error was less than half that of their validation error. Although this showed reliable performance in our experiments, when the risk of overfitting is high, it is arguably worthwhile validating, at least briefly, the quantile and coverage accuracy during model selection.

In a related note, it is important to utilise out-of-bag estimates for the mean function on the training set when using a bagged estimate for the mean, such as the random forest, as these provide a closer alignment in the magnitude of the residuals with those in unseen test cases.

Finally, the potential limitation of the proposed approach illustrated in scenario (ii) in the example in Section 2, in the context of vastly different residual scale, indicates that prediction intervals can be inaccurate over regions of data sparsity if the assumption on which our approach is based is violated to a large degree. This poses a restriction in applications involving risk, for example, where extremal quantiles associated with relatively rare events are of the utmost importance. While we have provided evidence that the proposal represents a useful general purpose tool for enhancing regression analysis for standard problems in which prediction accuracy and uncertainty quantification are of relatively equal importance for all points, further investigation is needed in order to determine its usefulness in some application areas.